\documentclass[fleqn,usenatbib]{mnras}

\usepackage{newtxtext,newtxmath}

\usepackage[T1]{fontenc}

\DeclareRobustCommand{\VAN}[3]{#2}
\let\VANthebibliography\thebibliography
\def\thebibliography{\DeclareRobustCommand{\VAN}[3]{##3}\VANthebibliography}

\usepackage{graphicx}	
\usepackage{amsmath}	
\usepackage{placeins}
\usepackage{subcaption}

\def\HI{\ion{H}{I}\,}
\def\xb{\bar{x}_{\rm \ion{H}{I}}}

\title[Inference using power spectrum and bispectrum]{Constraining the neutral hydrogen fraction during reionization: Cross-simulation inference using power spectrum and bispectrum}

\author[Krishna et al.]{
Anoop Krishna,$^{1}$\thanks{E-mail:krishnaanoop71@gmail.com}
Deepthi Moorkanat,$^{1}$
Hiten,$^{1}$
Rajesh Mondal$^{1}$
\\
$^{1}$Department of Physics, National Institute of Technology Calicut, Calicut 673601, Kerala, India\\
}

\date{Accepted XXX. Received YYY; in original form ZZZ}

\pubyear{\the\year{}}

\begin{document}
\label{firstpage}
\pagerange{\pageref{firstpage}--\pageref{lastpage}}
\maketitle

\begin{abstract}
The redshifted 21-cm signal is a unique probe of the early universe, particularly the Epoch of Reionization (EoR). While the 21-cm power spectrum has been the primary statistic for parameter inference, it fails to capture the non-Gaussian information in the signal, motivating the use of higher-order statistics such as the bispectrum. We perform a rigorous cross-simulation validation to infer the mean neutral hydrogen fraction ($\xb$) by training a neural network on {\sf 21cmFAST} simulations and applying it to mock observations generated by the {\sf ReionYuga} code. We first benchmark the framework in an idealized {\sf 21cmFAST}-only setting before applying it to the more rigorous {\sf ReionYuga}--{\sf 21cmFAST} cross-simulation case. Our analysis spans six redshifts and includes realistic SKA system noise and cosmic variance, calculated from 50 statistically independent realizations. In the same-code case, the bispectrum yields substantially tighter constraints, whereas in the cross-simulation case the improvement is moderate, with constraints tightened by $\sim 1.4\times$ relative to the power spectrum-only case. The cross-simulation analysis also identifies a persistent systematic discrepancy between inferred and true values that often exceeds the statistical uncertainties, implying that modeling uncertainty remains the dominant limitation. Our results, therefore, indicate that the highly stringent constraints obtained in same-code validation studies may be overly optimistic, and mitigating cross-model systematics is crucial for robust parameter inference in the SKA era.
\end{abstract}




\section{Introduction}
The Epoch of Reionization (EoR) marks a crucial phase transition in cosmic history, occurring roughly between redshifts $z\sim6-15$ \citep{Becker2001,Planck2020}. During this period, the first luminous sources gradually heated and ionized the neutral hydrogen (\ion{H}{I}) in the intergalactic medium (IGM) \citep{Furlanetto2006, Pritchard2012}. Probing this era directly remains one of the most ambitious goals in modern cosmology. The redshifted 21-cm emission from the hyperfine transition of \HI provides a direct probe of the reionization process in three dimensions \citep{Furlanetto2006, Morales2010}. Ongoing observational efforts from pathfinders such as the Low-Frequency Array (LOFAR), the Murchison Widefield Array (MWA), and the Hydrogen Epoch of Reionization Array (HERA) have now made statistical measurements of the EoR 21-cm signal possible (see, e.g., \citealt{Barry_2019, Abdurashidova_2023, Mertens_2025}).

The primary statistical measure for analyzing the 21-cm signal has been the power spectrum $P(k)$. The $P(k)$ captures the variance of brightness temperature fluctuations across different spatial scales. The $P(k)$ is commonly expressed in terms of the dimensionless form $\Delta^2(k)=k^3P(k)/2\pi^2$. LOFAR has reported upper limits on the EoR 21-cm power spectrum, $\Delta^{2} \lesssim (54.3 - 68.7\,\mathrm{mK})^{2}$ at a characteristic wavenumber of $k \approx 0.08\,h\,\mathrm{cMpc}^{-1}$ across the redshift range $z \sim 8.3-10.1$  \citep{Mertens_2020, Mertens_2025}. Meanwhile, MWA has provided deep limits of $\Delta^{2} \le (30.2\,\mathrm{mK})^{2}$ at $z = 6.5$ ($k = 0.18\,h\,\mathrm{Mpc}^{-1}$), along with broader constraints extending up to $\approx (62.4\,\mathrm{mK})^{2}$ across $z \sim 6.5 - 8.7$  \citep{Barry_2019, Li_2019, Trott_2020, nunhokee2025limits21cmpower}. The HERA Collaboration has reported improved upper limits to date, achieving $\Delta^{2} \lesssim (21.4 - 59.1\,\mathrm{mK})^{2}$ across $z \sim 8-10$ at $k \sim 0.35\,h\,\mathrm{Mpc}^{-1}$ \citep{Abdurashidova_2023,theheracollaboration2025resultsheraphaseii}. These limits disfavor extreme cold-reionization scenarios \citep{Ghara2020, Mondal2020b, Abdurashidova_2023}. Together, these developments mark the transition of this field from early instrument characterization to the era of active, quantitative EoR inference. The forthcoming SKA-Low is expected to further improve sensitivity and enable precision measurements of the 21-cm signal.

The power spectrum is fundamentally limited when dealing with non-Gaussian fields. The EoR 21-cm signal is highly non-Gaussian \citep{Bharadwaj2004, Bharadwaj2005, Mondal2015} due to the underlying non-linearities in the matter field and the complex, inhomogeneous nature of the reionization process. As ionized bubbles form around sources and grow until they eventually percolate \citep{Iliev2006}, they introduce complex phase correlations that the power spectrum, being second-order statistics, cannot resolve. Higher-order statistics, starting with the bispectrum, are therefore needed to fully describe the reionization process and break degeneracies between different physical drivers \citep{Watkinson2017, Majumdar2018, Mondal2020}.

The current challenge is to constrain the parameters that provide information regarding the underlying IGM and astrophysics. This approach typically involves Bayesian inference, often implemented via Markov Chain Monte Carlo (MCMC) sampling. Such inference is now computationally feasible due to semi-numerical simulation tools such as {\sf ReionYuga}~\citep{Mondal2017} and {\sf 21cmFAST}~\citep{Mesinger2011}, as well as other variants, often coupled with neural network emulators for rapid likelihood evaluation. 

Recent work by \citet{Tiwari2022} demonstrated that including the bispectrum in parameter inference significantly enhances the constraints on EoR parameters. However, the study was limited in two key respects. Firstly, it relied on `same-model' validation. When the same simulation framework is used for both emulator training and the generation of the fiducial `observed' data, the results may be artificially precise and fail to account for the systematic modeling uncertainties. Second, the study shares a common vulnerability in its error treatment, utilizing sample variance under the Gaussian assumption for a signal that is known to be highly non-Gaussian. This is self-contradictory and underestimates the true errors \citep{Mondal2016, Mondal2021}. This also yields biased estimates of the inferred parameters \citep{Shaw2020, Mondal2022}. Alternate approaches such as simulation-based inference (SBI) offer a framework for combining multiple summary statistics, including bispectrum and topological descriptors, while naturally accommodating intractable likelihoods and providing greater flexibility \citep{Cerardi2025}.

In this paper, we revisit these problems by performing a cross-simulation validation while properly accounting for signal non-Gaussianities in the error budget. Our analysis incorporates a realistic treatment of the SKA-Low error budget, including non-Gaussian cosmic-variance errors and system noise. We use a Neural Network emulator trained on the {\sf 21cmFAST} simulation data to infer the mean neutral hydrogen fraction ($\xb$) from mock observations generated by the {\sf ReionYuga} simulation. Because astrophysical parameters are model-dependent and carry different physical meanings across different simulation codes, we focus on $\xb$ as a direct observable quantity independent of the specific model. We demonstrate that while the bispectrum provides a clear statistical gain, the improvement is not drastic, as previously reported \citep{Tiwari2022} when more realistic error treatments are applied. Furthermore, we examine the systematic offsets between different simulation approaches, which pose a major challenge for inference. The same has been observed in SKA SDC3b: EoR inference\footnote{\url{https://sdc3.skao.int/challenges/inference}}, suggesting that future constraints will be limited by modeling accuracy. 

Throughout the paper, we adopt the Planck+WP best-fitting values of cosmological parameters \citep{Planck2020}.


\section{Simulating the fiducial model}
We generate the `true' or fiducial observations using the semi-numerical code {\sf ReionYuga}\footnote{\url{https://github.com/rajeshmondal18/ReionYuga}}~\citep{Mondal2017}. This mock observation is intentionally different from the training dataset used for the emulator to ensure a rigorous test of cross-code validity. The simulation process follows three steps. First, it employs a particle-mesh (PM) $N$-body code\footnote{\url{https://github.com/rajeshmondal18/N-body}}~\citep{Mondal2015} to evolve the dark matter distribution. Second, the collapsed halos, which serve as hosts for the ionizing sources, are identified using a Friends-of-Friends (FoF) halo finder\footnote{\url{https://github.com/rajeshmondal18/FoF-Halo-finder}}~\citep{Mondal2016}. Finally, the ionizing field is constructed based on the set excursion formalism \citep{Furlanetto2004}.

Our fiducial ionization simulation is performed in a cubic comoving volume with a side length of $215.04\,{\rm Mpc}$ and spatial resolution of $0.56\,{\rm Mpc}$. The specific model parameters and simulation setup used here are taken directly from \citep{Mondal2017}. For an exhaustive description of the simulation physics and the parameter values, we refer the reader to that work. We analyze six snapshots at redshifts $z = \{7.0,8.0,9.0,10.0,11.0,13.0\}$. At each redshift, we use 50 statistically independent realizations, allowing us to estimate the mean power spectrum and bispectrum, as well as associated cosmic variance errors.

The total uncertainty in our mock measurements is driven by two primary components, system noise ($\sigma_{\rm sys}$) and cosmic variance ($\sigma_{\rm CV}$). System noise is an instrumental effect that is statistically uncorrelated between modes and dominates the error budget at small spatial scales. Conversely, cosmic variance represents the intrinsic statistical uncertainty arising from the finite cosmological volume accessible to observers, dominating on large scales due to the limited number of independent Fourier modes available (see e.g., \citealt{Mondal2015}).

Rather than adopting a simplified analytical model for these errors, we use a simulation-based approach to calculate the total noise (cosmic variance + system noise). System noise for a 128\,hour observation with the future SKA-Low is generated following the interferometric noise methodology detailed in \citet{Shaw2019}. We inject 50 independent realizations of this thermal noise directly into the 50 brightness temperature ($\delta T_{\rm b}$) cubes. Subsequently, the total errors in the power spectrum and bispectrum measurements are calculated from these 50 statistically independent realizations of the signal. This ensures that our estimates naturally incorporate the non-Gaussian nature of the 21-cm signal. For further discussion of this error estimation method, the reader is referred to \citet{Mondal2021}. We also check the convergence of the error estimates as a function of the number of realizations that are presented in Appendix~\ref{app:variance_convergence}. We note that the power spectrum variance is largely stabilized, while the bispectrum variance is approaching convergence at $N=50$ and is adequate for the present analysis.

\section{Simulating the Inference Model}
\label{sec:inference_model}
To simulate inference models, we use the publicly available {\sf 21cmFAST} code \citep{Mesinger2011}. This framework efficiently produces three-dimensional brightness temperature fields by applying first-order perturbation theory for structure formation and an excursion-set formalism to identify ionized regions based on the local collapsed fraction. To maintain consistency, we simulate a box size and spatial resolution identical to those used in our fiducial model. 

The ionization process is parameterized mainly using three physically motivated quantities. First, the ionizing efficiency parameter {\sf HII\_EFF\_FACTOR} (or $\zeta_0$), which represents a combination of several factors, e.g., the star formation efficiency, the number of ionizing photons produced per baryon, and the escape fraction of these photons. Second, {\sf R\_BUBBLE\_MAX} defines the maximum mean free path of ionizing photons. Third, {\sf ION\_Tvir\_MIN} represents the minimum virial temperature required for a halo to host star-forming galaxies \citep{Greig2015}. We generate $\delta T_{\rm b}$ snapshots at the same six redshifts used for the fiducial model. The three-parameter set $\{\zeta_0,\, {\sf R\_BUBBLE\_MAX},\, {\sf ION\_Tvir\_MIN}\}$ adopted here constitutes a standard physically motivated minimal parameterization of EoR astrophysics with sufficient flexibility to reproduce a broad class of reionization histories and enables straightforward interpretation \citep{Greig2015}. However, this compact parameterization also imposes structural limitations on the {\sf 21cmFAST} model space and may therefore not capture the full range of physically plausible reionization scenarios.

A dataset was constructed using Latin hypercube sampling with 1000 points to ensure efficient coverage of the parameter space. The parameter ranges span ${\sf HII\_EFF} \in [10,100]$, ${\sf R\_BUBBLE\_MAX} \in [5,30]$ and $\log_{10}({\sf ION\_Tvir\_MIN}/{\rm K}) \in [4.0,5.7]$. To examine whether the {\sf 21cmFAST} training set provides sufficient coverage of the {\sf ReionYuga} ionization history across redshift, we directly compare the corresponding ionization histories in Appendix~\ref{app:xhi_comparison}. We find that the {\sf ReionYuga} ionization history is shifted toward slightly later reionization relative to the median of the {\sf 21cmFAST} model ensemble. However, the true value of $\xb$ at every redshift considered is contained within the range spanned by the training set. This implies that the emulator is applied strictly within its training domain at all analyzed redshifts, so that the inference is not driven by extrapolation. From each simulation box, we extract the $\xb$ and compute summary statistics, i.e., the power spectrum and the bispectrum. In our emulator, discussed in Section~\ref{sec:emulators}, these $\xb$ serve as input while the summary statistics are the output.


\section{Summary Statistics}
We characterize the 21-cm signal using two summary statistics: the spherically averaged power spectrum (SAPS) and the spherically averaged bispectrum (SABS). Together, these statistics capture both the Gaussian and the higher-order non-Gaussian information content of the brightness temperature field.

\begin{figure*}
\centering
\includegraphics[width=\textwidth]{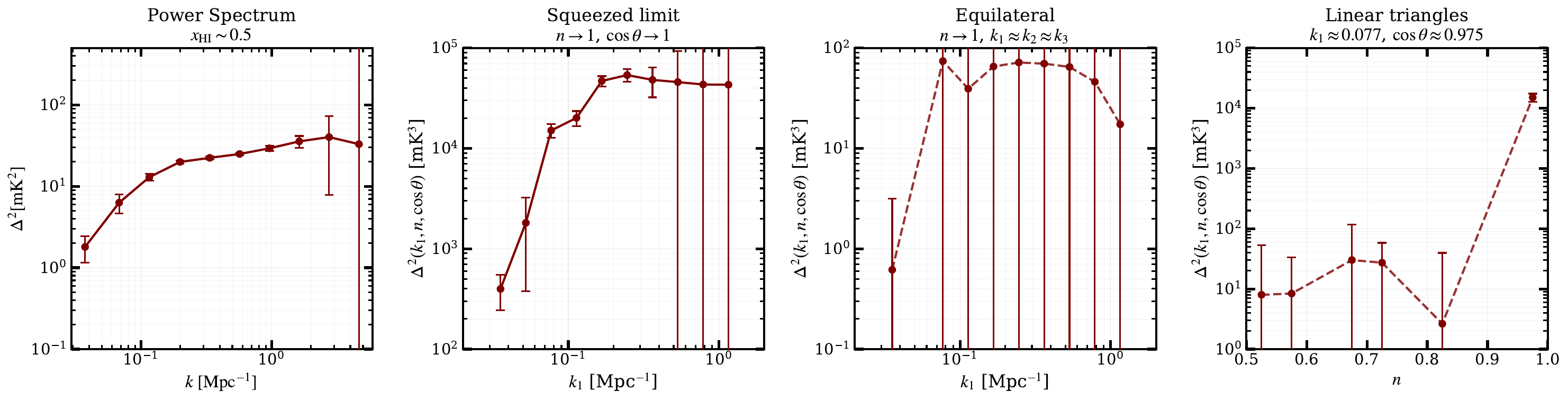}
\caption{Summary statistics of the fiducial 21-cm signal at $z=8$. The panels display (from left to right): the SAPS, followed by the SABS in the squeezed limit, equilateral configuration, and linear triangles.}
\label{fig:combined_stats}
\end{figure*}

The power spectrum quantifies the spatial fluctuations in $\delta T_{\rm b}$ at various length scales. Assuming statistical homogeneity, it is defined as \citep{Mondal2015}
\begin{equation}
P(\mathbf{k}) = V^{-1}\langle\Delta_{\rm b}(\mathbf{k})\Delta_{\rm b}(-\mathbf{k})\rangle \, ,
\end{equation}
where $\Delta_b(\mathbf{k})$ is the Fourier transform of $\delta T_b$ and $V$ is the comoving volume under consideration. The dimensionless power spectrum for the fiducial model at $z=8$ is shown in the left panel of Figure~\ref{fig:combined_stats}. The power spectrum increases monotonically toward smaller scales (larger $k$), characteristic of the mid-reionization regime, where the growth of ionized bubbles and the density contrast on these scales enhance the power \citep{Bharadwaj2004, Mondal2017}. On larger scales (small $k$), fluctuations are relatively suppressed because the brightness temperature distribution is more uniform. In the figure, we also show the 1-$\sigma$ errors ($\sigma_{\rm sys} + \sigma_{\rm CV}$) on the power spectrum calculated from the ensemble of 50 statistically independent realizations of the fiducial signal.

The bispectrum is the lowest-order statistic that can capture the non-Gaussianity in a signal. It records the correlation between three Fourier modes and is defined as \citep{Bharadwaj2005, Mondal2021}
\begin{equation}
B(\mathbf{k}_1,\mathbf{k}_2,\mathbf{k}_3) = V^{-1} \delta^{\rm K}_{\mathbf{k}_1 +\mathbf{k}_2+\mathbf{k}_3} \langle\Delta_b(\mathbf{k}_1)\Delta_b(\mathbf{k}_2)\Delta_b(\mathbf{k}_3)\rangle \, .
 \end{equation} 
Here, the Kronecker delta function $\delta^{\rm K}$ imposes the triangular closure condition $\mathbf{k}_1+\mathbf{k}_2+\mathbf{k}_3=0$. We compute the SABS for all unique closed triangle configurations using the \textsc{DviSukta} code\footnote{\url{https://github.com/rajeshmondal18/DviSukta}} \citep{Mondal2021}, employing the $(k_1,\,n,\,\cos\theta)$ parametrization from \citet{Bharadwaj2020}. Here, $k_1=|\mathbf{k}_1|$ is the longest side, $n=k_2/k_1$ ($\in(0,1]$) is the side ratio, and $\cos\theta=-\hat{\mathbf{k}}_1\cdot\hat{\mathbf{k}}_2$ ($\in [0.5,1]$) defines the shape of the triangle. In Figure~\ref{fig:combined_stats}, we also show the dimensionless SABS, $\Delta^3(k_1,n,\cos\theta)$ for the fiducial model at $z=8$ for the squeezed limit ($n\to1$, $\cos\theta\to1$, so $k_3\to0$), equilateral triangles ($n\to1$, $\cos\theta\to0.5$, so $k_1\approx k_2\approx k_3$), and linear (collinear) triangles ($\cos\theta\to1$ fixed, $n$ varying). The squeezed-limit bispectrum typically shows the largest amplitude, reflecting the strong coupling between large-scale modulation and small-scale ionization structures \citep{Majumdar2018}. Other shapes capture important EoR physics because different triangle configurations probe different mode couplings and hence distinct aspects of the signal \citep{Mondal2021}. The 1-$\sigma$ errors ($\sigma_{\rm sys} + \sigma_{\rm CV}$) on the SABS were calculated from the fiducial signal ensemble. In our analysis, we retain only configurations with signal-to-noise ratio $\mathrm{SNR}>1$.

\section{Emulators}
\label{sec:emulators}
To facilitate rapid exploration of the parameter space, we construct neural network emulators for both the power spectrum and bispectrum at $z =\{7,8,9,10,11,13\}$. We used the Keras package in TensorFlow \citep{tensorflow2015-whitepaper,chollet2015keras} to construct the networks.

For the power spectrum, we train a separate Bayesian neural network (BNN) at each redshift to emulate the mapping $\bar{x}_{\mathrm{H,I}} \rightarrow \Delta^2(k)$ on a fixed $k$-grid. Each network comprises three fully connected hidden layers with 64, 128, and 64 neurons, respectively, with Rectified Linear Unit (ReLU) activations. To capture epistemic uncertainty, we employ Monte Carlo dropout \citep{Gal2016} with a dropout rate of $p = 0.10$, retained during both training and inference. The mean and variance over repeated stochastic forward passes then define the emulator prediction, $\bar{\Delta^2}(k)$, and the corresponding epistemic uncertainty, $\sigma_{\mathrm{emu}}^{P}$. This is also consistent with the recent BNN-based inference study of \citet{Mahida2025}, who showed that Bayesian emulators can produce more robust constraints than deterministic ANN emulators when the uncertainty of the emulator is consistently propagated. The emulator achieves a test-set coefficient of determination of $R^{2} = 0.98$.

The bispectrum presents a significantly more demanding emulation problem than the power spectrum. Its amplitude exhibits substantial variation across triangle configurations and redshifts, and may also change sign. We therefore adopt a \emph{pointwise ensemble emulator} that learns the scalar mapping $(\xb, \alpha) \rightarrow \Delta^{3}(\alpha)$, hereafter denoted as $b$, where $\alpha \equiv (k_{1},n,\cos\theta)$. In this formulation, the triangle configuration enters directly as part of the input space, which provides a more flexible representation of the bispectrum across different configurations and redshifts. To handle the sign changes in $b$, we decompose the target as $b = \mathrm{sgn}(b)\,|b|$ and adopt a two-headed network architecture. The shared trunk comprises three fully connected layers with 128, 128, and 64 neurons, using ReLU activations and $\ell_{2}$ weight regularization. It then bifurcates into a magnitude head, which predicts $\log(1+|b|)$ through a softplus activation and is trained with a mean-squared-error loss, and a sign head, which predicts the probability of $b \geq 0$ through a sigmoid activation and is trained with binary cross-entropy. The total loss is defined as the equally weighted sum of these two components. 

We estimate the epistemic uncertainty in the bispectrum emulator using an ensemble of $N_{\mathrm{ens}} = 4$ independently trained networks \citep{lakshminarayanan2017}. The ensemble mean provides the point prediction $\bar{b}$, while the ensemble standard deviation, $\sigma_{\mathrm{emu}}^{B}$, is included in the bispectrum likelihood as an additional variance term. To prevent information leakage, the 1,000 simulations are partitioned at the level of astrophysical parameter groups, so that all triangle configurations from a given simulation remain within the same split. The data grouped are divided into training and test sets in a 90/10 ratio.

The ensemble bispectrum emulator achieves a mean test-set $R^{2}$ of 0.97. Further diagnostic plots, showing the predicted and true bispectrum at $z = 8$ for configurations with SNR $> 1$ are included in Appendix~\ref{app:emulator_performance}. 

The ensemble predictions track the 1:1 line tightly across more than five orders of magnitude in amplitude, confirming reliable emulation of the dominant non-Gaussian configurations. The overall ensemble test $R^2$ ranges from $0.952$ at $z = 7$ to $0.997$ at $z = 13$.

\section{Inference framework}
\label{sec:inference}
We perform Bayesian parameter inference using the {\sf emcee} ensemble sampler \citep{Foreman-Mackey2013}. Our analysis compares two cases: (i) an inference based exclusively on power spectrum, and (ii) a joint analysis combining $P(k)$ and $B(k_1, n, \cos \theta)$. To ensure that $\xb \in [0,1]$, we operate in logit-transformed space, $u_z = \mathrm{logit}[\xb(z)]$. 

A central challenge in cross-simulation validation is the inherent systematic bias between different modeling frameworks. To address this, we adopt an input calibration scheme motivated by the Bayesian calibration formalism for computer models developed by \citep{Kennedy2002} and extended by \citep{Higdon01062008}. The emulator is trained on a globally calibrated input, $\xb^{\rm emu} = \sigma(u_z + \delta_{\rm global})$, where $\sigma$ is the sigmoid function and $\delta_{\rm global}$ is the input deformation factor that accounts for the systematic bias in the emulator. Additionally, a redshift-dependent amplitude calibration $A(z) = A_1 (z - z_\mathrm{ref}) + A_0$ is utilized to capture multiplicative offsets, where $z_\mathrm{ref} = 10$ is chosen as the pivot redshift. This linear parametrization allows the calibration to adapt to redshift-dependent differences in the ionization history (Figure~\ref{fig:xhi_history}b) while maintaining computational tractability. We apply uniform priors $A_1 \in [-1.0, 1.0]$ and $A_0 \in [-5.0, 5.0]$ 
directly to the linear amplitude calibration $A(z)=A_1(z-z_{\rm ref})+A_0$, requiring $A(z)>0$ at all analysed redshifts, and $\delta_\mathrm{global} \in [-2,2]$ to the logit-space shift in $\xb$. By marginalizing over the nuisance parameters $\mathbf{A_1,A_0}$ and $\delta$, we obtain meaningful constraints on $\xb$ while mitigating the effects of cross-simulation mismatches. Appendix~\ref{app:calibration_impact} shows that without these calibration parameters, the cross-simulation inference catastrophically fails at $z = 7$--$9$. This strategy will be vital for future SKA observations (see e.g., SKA SDC3b: EoR inference\footnote{\url{https://sdc3.skao.int/challenges/inference}}).

\begin{figure*}
    \centering
    \includegraphics[width=\textwidth]{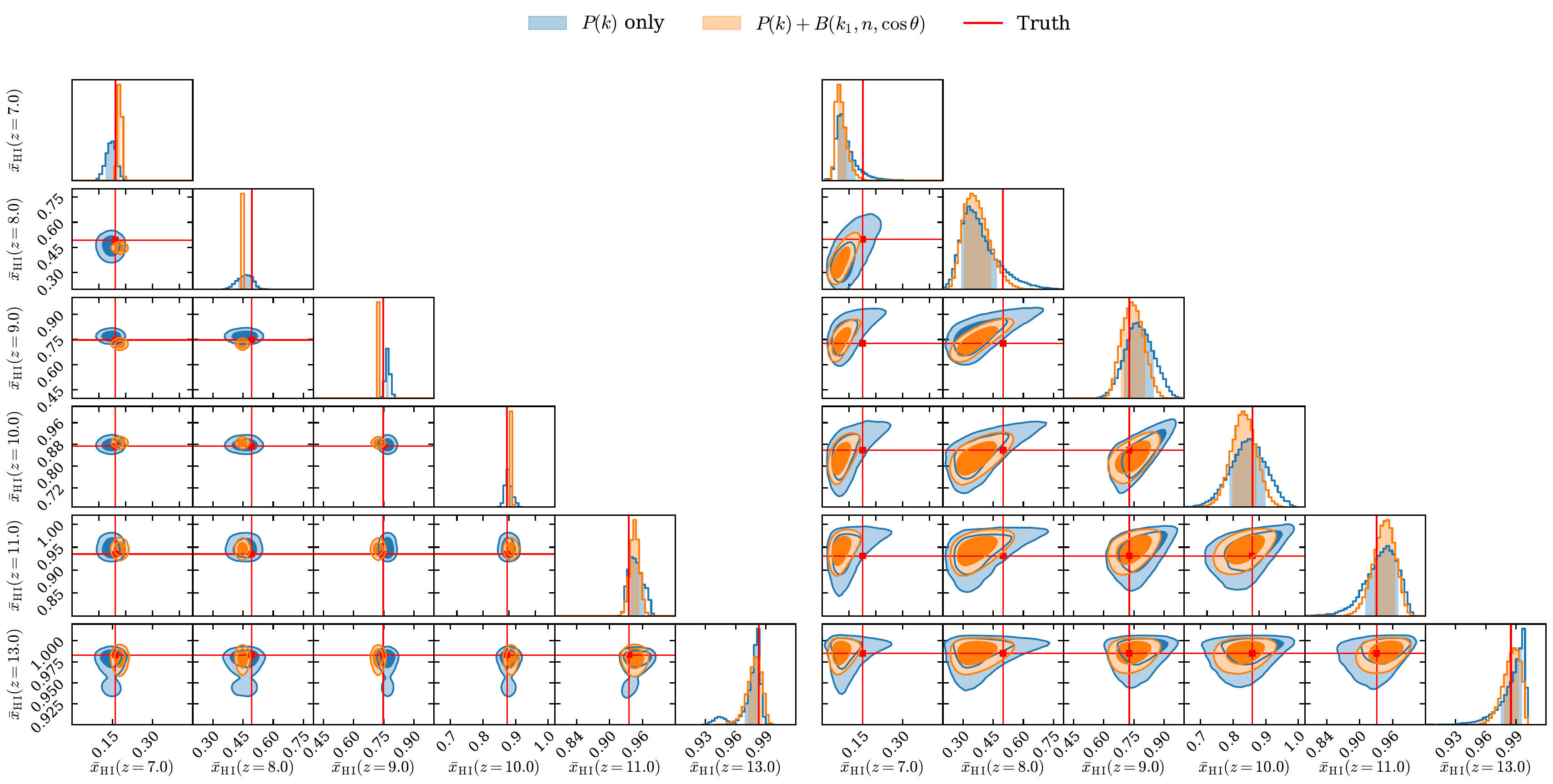}
    \caption{Posterior distributions (1D and 2D) of $\xb$ at $z=\{7,8,9,10,11,13\}$ for the power spectrum-only analysis (blue) and the joint power spectrum--bispectrum analysis (orange). The left panel corresponds to the same-code validation using {\sf 21cmFAST} for both the mock input and the inference model, while right panel shows the cross-simulation case with {\sf ReionYuga} mock input and {\sf 21cmFAST}-based inference. The shaded contours denote the $1\sigma$ and $2\sigma$ credible regions, and the red lines indicate the true values.}
    \label{fig:posterior}
\end{figure*}

For the power spectrum, the likelihood is defined as: 
\begin{equation}
\label{eq:lh}
\mathcal{L}_{\rm P} \propto \exp\left[ -\frac{1}{2} \left( \boldsymbol{\Delta}_{\rm P}^T \mathbf{C}_{\rm P}^{-1} \boldsymbol{\Delta}_{\rm P} \right) \right],
\end{equation}
where the residual vector $\boldsymbol{\Delta}_{\rm P} = P_{\rm obs}(k) - A \cdot P_{\rm emu}(k \mid \xb^{\rm emu})$. The covariance matrix $\mathbf{C}_{\rm P}$ 
incorporates cosmic variance, system noise, epistemic uncertainty, and a conservative 20\% modeling error $\sigma_{\rm model}$. We tested the sensitivity of our results to this assumption by increasing the bispectrum modeling error term to 40\%. While this broadens the posterior credible intervals, as expected, it does not change the posterior centres or the qualitative conclusions. This indicates that the residual offset observed in the cross-simulation case is driven primarily by structural model mismatch (Figure~\ref{fig:posterior}, right panel), rather than by an underestimation of the adopted diagonal error budget.

For the bispectrum, we employ a likelihood formalism similar to equation \ref{eq:lh}. However, because computing the full covariance for hundreds of triangle configurations is computationally expensive, we adopt a diagonal likelihood approximation. We replace the full covariance matrix with the diagonal variance $\sigma_{B}^{2}(z,\alpha$), where $\alpha \equiv (k_{1},n,\cos\theta)$. This approach captures the dominant contributions to the likelihood and is commonly adopted in previous bispectrum analyses \citep{Mondal2015, Shaw2020}.
For the combined analysis, we employ an informative multivariate Gaussian prior on $\xb(z)$, obtained from the power spectrum MCMC posterior, acknowledging that any bispectrum measurement would be determined in combination with existing power spectrum constraints. The MCMC sampling is performed with $N_{\rm walkers} = 64$, running for up to 30,000 steps, where adaptive convergence is implemented.
To evaluate the impact of cross-simulation systematics, we first validate the inference framework in an idealized same-code space, where both the fiducial mock observation and the emulator training set are from {\sf 21cmFAST}. In this baseline case, no additional calibration parameters are introduced. We choose the fiducial {\sf 21cmFAST} realization such that its reionization history most closely matches that of the {\sf ReionYuga} fiducial. This ensures that the comparison is not driven primarily by differences in the underlying ionization history.



\section{Results}
\begin{figure*}
    \centering
    \includegraphics[width=\textwidth]{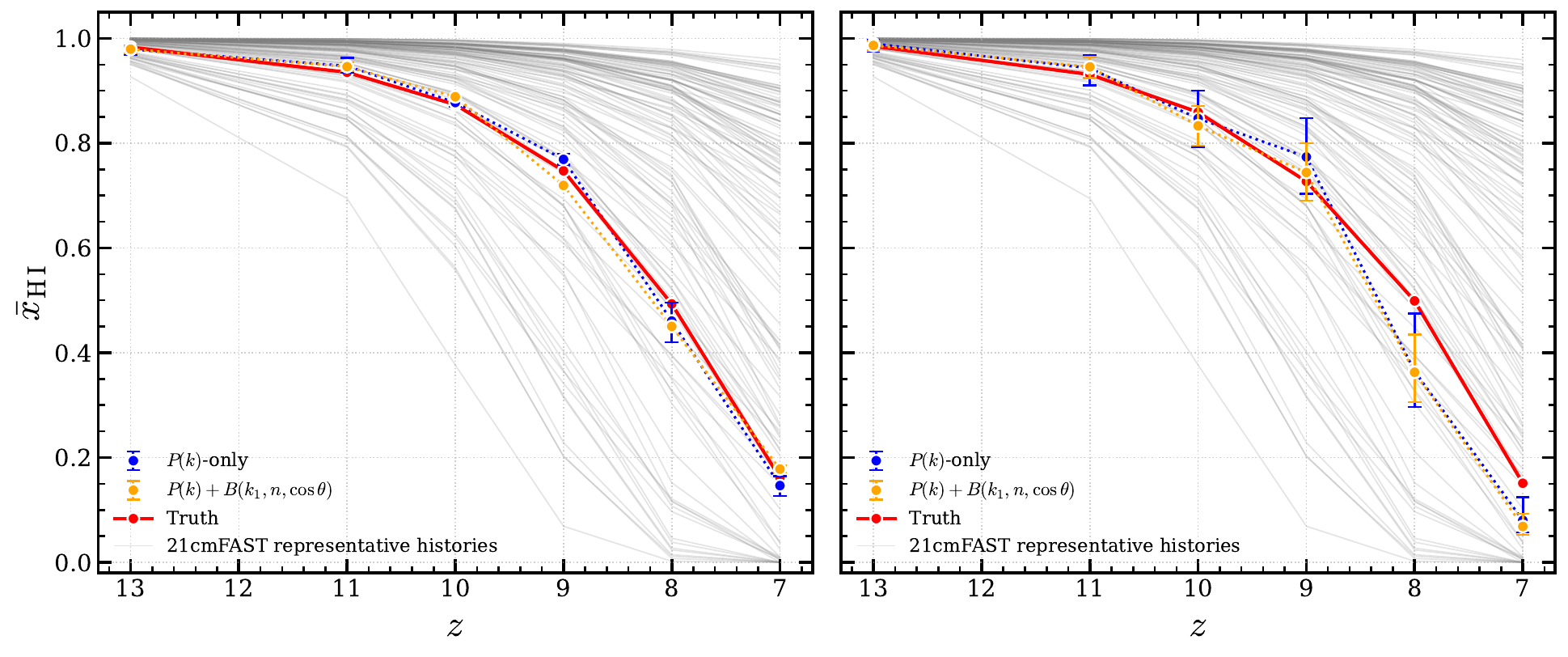}
    \caption{Inferred ionization histories projected onto the $\xb$--$z$ plane for the same-code and cross-simulation cases. The left panel shows the same-code case, and the right panel shows the cross-simulation case. Grey lines show representative ionization histories from the {\sf 21cmFAST} training set. The red line with filled circles denotes the fiducial truth. Blue and orange points with error bars show the inferred $\xb$ from the power spectrum alone and from the joint power spectrum--bispectrum analysis, respectively.}
    \label{fig:xhi_history}
\end{figure*}

We first validate the inference framework in an idealized same-code setting, in which both the fiducial observation and the emulator are based on {\sf 21cmFAST}. The resulting posteriors for $\xb (z)$ at the six analysis redshifts are shown in Figure~\ref{fig:posterior} (left panel). We quantify the gain from adding the bispectrum using the ratio of the posterior 68\% credible-interval widths,
\begin{equation}
    \mathcal{S}(z) =
    \frac{W^{P}_{68}(z)}
         {W^{P+B}_{68}(z)} ,
\end{equation}
where \(W_{68}=q_{84}-q_{16}\). Thus, \(\mathcal{S}>1\) indicates that the joint \(P(k)+B(k_1,n,\cos\theta)\) analysis gives a tighter constraint than the power spectrum-only case. The power spectrum alone (blue contours) yields tight constraints, with typical $1\sigma$ fractional uncertainties of ${\sim}5$--$10$ per cent.  The joint analysis including the bispectrum (orange contours) further sharpens these constraints, producing posteriors that are approximately $16\times$ narrower (\(\mathcal{S}=16\)) than in the power spectrum-only case. This demonstrates the additional constraint provided by the non-Gaussian information encoded in the bispectrum when no simulation-dependent model mismatch is present. Figure~\ref{fig:xhi_history} (left panel) presents these constraints in the $\xb$--$z$ plane, overlaid on representative ionization histories drawn from the {\sf 21cmFAST} training set (grey lines). The inferred constraints follow the {\sf 21cmFAST} truth (red line) closely at all redshifts, with only small residual offsets. The offsets in the same-code case can be attributed to residual degeneracies in the $\xb$ summary statistic mapping, since different ionization morphologies can produce similar power spectrum and bispectrum amplitudes at fixed redshift.

The right panel of Figure~\ref{fig:posterior} shows the posterior corresponding to the cross-simulation scenario, where the mock observation is generated by {\sf ReionYuga} while the emulator remains trained on {\sf 21cmFAST}. The results for $P(k)$ alone are shown in blue, and for $B(k_1, n, \cos \theta)$ with informative prior from $P(k)$ are shown in orange. The joint analysis yields constraints on $\xb$ that are roughly $1.4$ times more stringent than the power spectrum results. Notably, the degree of improvement does not scale linearly with the number of triangle configurations. For instance, constraints at $z=9$ and $z=10$ show better fractional gains than at $z=8$, despite having fewer triangle configurations (${\rm SNR}>1$). This reflects the fact that the bispectrum provides non-redundant information that is unique to specific stages of reionization. Figure~\ref{fig:xhi_history} (right panel) shows the corresponding projection of the ionization history for the cross-simulation case. In contrast to the same-code case, the inferred $\xb$ values show systematic offsets relative to the {\sf ReionYuga} truth (red line).

The redshift-dependent calibration slightly reduces systematic offsets relative to a global-$A$ treatment (see Appendix~\ref{app:calibration_impact}), though residual biases remain visible at lower redshifts. While $A$ and $\delta$ partially mitigate discrepancies in cross-simulation validation (see Appendix~\ref{app:calibration_impact} for a direct comparison), they cannot fully capture the redshift-dependent physical differences between {\sf ReionYuga} and {\sf 21cmFAST}. A fully redshift-dependent treatment of both calibration parameters would substantially increase the dimensionality of the MCMC inference, and make the resulting discrepancy model considerably harder to interpret. Therefore, the linear \(A(z)\) model reduces the cross-simulation bias relative to the global-calibration case, but the inferred histories still remain systematically offset from the {\sf ReionYuga} truth at the lowest redshifts. Thus, a linear redshift dependence improves the fit, but does not fully eliminate the underlying model mismatch. This persistent bias, consistent with the findings of \citet{Zahn2011, Majumdar2014}, suggests that future constraints may be more limited by the algorithmic assumptions of our simulation frameworks. 

These results reveal a clear distinction between the same-code and cross-simulation analysis. The same-code case yields substantially tighter constraints and a much stronger apparent gain from including the bispectrum, whereas the cross-simulation case results in broader posteriors, only a modest improvement from the bispectrum, and visible residual offsets between the inferred and true $\xb$ values. This contrast is evident both in Figure~\ref{fig:posterior} and in the corresponding $\xb$--$z$ evolution shown in Figure~\ref{fig:xhi_history}.


\section{Conclusion \& Discussion}
We have performed a stringent cross-simulation validation of 21-cm inference during the EoR, training a NN on {\sf 21cmFAST} simulation and applying it to a `true' model generated by the independent {\sf ReionYuga} framework. This setup ensures that the interpolation capability of emulators is tested against an independent and potentially more realistic model of the universe. Our analysis jointly utilizes the power spectrum and the bispectrum to constrain $\xb$ at six different redshifts $z=\{7,8,9,10,11,13\}$. The error budget rigorously incorporates cosmic variance, system noise, $\xb$-dependent epistemic uncertainty, and modelling error. Notably, the cosmic variance and system noise are accounted for directly at the level of the 21-cm brightness-temperature cubes. Thereafter, all statistics are computed from 50 statistically independent realizations of the signal. This treatment is more robust than the Gaussian approximations common in previous studies, which underestimate uncertainties by 20–50\% and yield biased estimates of the inferred parameters \citep{Trott2016, Mondal2016, Greig2017, Shaw2020, Tiwari2022}.

Rather than adopting model-dependent astrophysical parameters, we infer a direct observable $\xb$. To isolate the impact of cross-simulation systematics, we first validate the framework in an idealized same-code setting, in which both the mock observation and the emulator are based on {\sf 21cmFAST}, and then apply the same framework to the cross-simulation case, where the mock observation is generated using {\sf ReionYuga}. In the same-code validation, the bispectrum tightens constraints by ${\sim}16\times$ relative to the power spectrum alone, with typical fractional uncertainties of 5--10 per cent. However, in the cross-simulation case, the bispectrum improvement is moderate (${\sim}1.4\times$) and posteriors are substantially wider. The improvement is consistent with the theoretical expectations that higher-order statistics encode additional information inaccessible to the power spectrum alone \citep{Watkinson2017, Majumdar2018, Mondal2020, Kamran2021}. In addition, we find that the information content does not scale linearly with the addition of the bispectrum of different configurations. Instead, a small number of effective configurations, such as the squeezed limit, capture the bulk non-Gaussian information in the signal. This is consistent with recent studies showing that not all bispectrum triangle configurations are equally informative, and that a subset of large-scale shapes, especially squeezed-limit and linear configurations, can be particularly effective in distinguishing different reionization scenarios \citep{Noble2024}.

Despite the inclusion of a redshift-dependent amplitude calibration together with a constant shift parameter, $\delta_\mathrm{global}$, residual systematic biases remain visible in the cross-simulation posteriors. The redshift-dependent treatment provides a modest improvement over the global-$A$ case, but does not fully remove the remaining offsets. This reflects the highly model-dependent nature of $\xb$ when derived from specific simulation frameworks. The bias remains evident in both the power spectrum and joint analysis, becoming even more evident when the bispectrum tightens the statistical bounds. A similar phenomenon is also observed in the SKA SDC3b. While the nuisance parameters $A$ and $\delta$ partially mitigated these biases, they are still inadequate to capture the redshift-dependent offset. This indicates that the discrepancy does not arise from simple normalization or calibration differences but instead reflects fundamental structural differences in how reionization physics is implemented across simulation frameworks.

The present {\sf 21cmFAST} training set adopts a restricted parametrization, in which the astrophysical parameters are treated as global, redshift-independent quantities. A more expansive parametrization, such as the mass- and redshift-dependent UV galaxy-luminosity prescription of \citet{Park2019} available in recent {\sf 21cmFAST} releases \citep{Murray2020}, would enlarge the family of reionization histories accessible to the inference model and could plausibly reduce the mismatch with {\sf ReionYuga} for some histories. However, this limitation is, to a large extent, precisely the effect that our cross-simulation analysis is designed to reveal. In any realistic application to future SKA data, the inference model will inevitably provide only a finite parametrized approximation to the underlying reionization physics, and an independent simulation or the real Universe itself may lie partly outside the model manifold. We therefore interpret the residual redshift-dependent bias found here not merely as a limitation of the specific three-parameter setup, but as a concrete manifestation of model-mismatch bias in 21-cm inference. Moreover, adopting a more flexible source model would generally increase the dimensionality of the inference problem, since the additional astrophysical freedom would need to be explored during MCMC. This would make the posterior structure more complex and the inference correspondingly more computationally demanding.

While increasingly flexible calibration schemes can mitigate cross-simulation systematics, they cannot fully solve the problem. More sophisticated approaches, such as emulator ensembles trained on multiple simulation codes, may be required for unbiased inference from future SKA observations \citep{Jasper2026}. A key implication for future SKA observations is that reports of sub-percent uncertainty on $\xb$ may be overconfident if they rely on `same-code' validation. Our cross-simulation results highlight that systematic modeling uncertainty remains the dominant bottleneck. Accurate inference will therefore require statistical frameworks that explicitly account for these model-dependent effects.

\section*{Acknowledgements}
RM is supported by the NITC FRG Seed Grant (NITC/PRJ/PHY/ 2024-25/FRG/12).
\begin{figure*}
\renewcommand{\thefigure}{A1}
\centering    
\includegraphics[width=\textwidth]{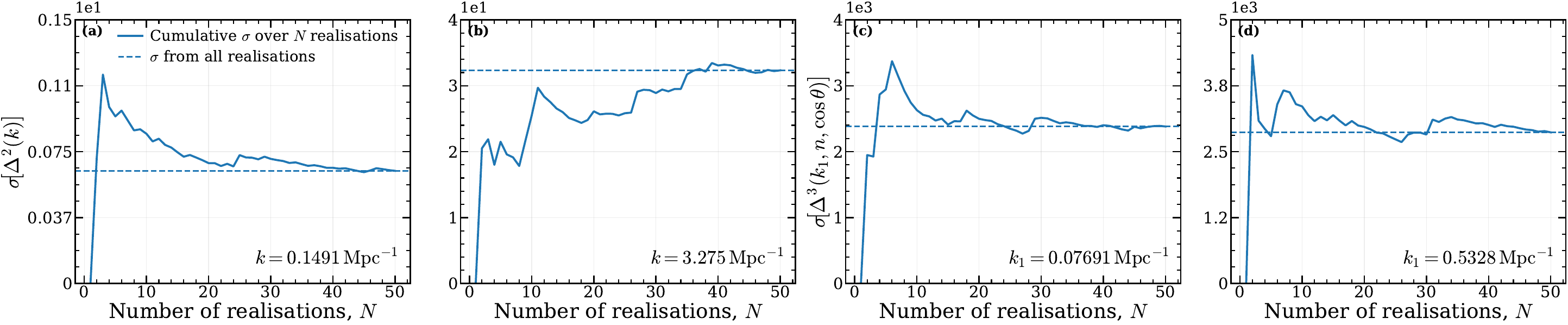}
\caption{Convergence diagnostic for the standard deviation estimated from the ensemble of realizations at $z = 8$. Panels (a) and (b) show the power spectrum for two representative $k$-modes. Panel (c) shows the bispectrum in the squeezed limit ($n \sim 1$, $\cos\theta \sim 1$), and panel (d) shows the bispectrum for equilateral configurations ($n \sim 1$, $k_{1} \sim k_{2} \sim k_{3}$). In each panel, the solid line denotes the cumulative standard deviation computed using the first $N$ realizations, while the dashed horizontal line marks the final value obtained from all 50 realizations.}
\label{fig:variance_convergence}
\end{figure*}
\section*{Data availability}
The data underlying this article will be shared on a reasonable request to the corresponding author.
\bibliographystyle{mnras}
\bibliography{bib} 

@misc{nunhokee2025limits21cmpower,
      title={Limits on the 21 cm power spectrum at z=6.5-7.0 from MWA observations}, 
      author={C. D. Nunhokee and D. Null and C. M Trott and N. Barry and Y. Qin and R. B. Wayth and J. L. B. Line and C. H. Jordan and B. Pindor and J. H. Cook and J. Bowman and A. Chokshi and J. Ducharme and K. Elder and Q. Guo and B. Hazelton and W. Hidayat and T. Ito and D. Jacobs and E. Jong and M. Kolopanis and T. Kunicki and E. Lilleskov and M. F. Morales and J. C. Pober and A. Selvaraj and R. Shi and K. Takahashi and S. J. Tingay and R. L. Webster and S. Yoshiura and Q. Zheng},
      year={2025},
      eprint={2505.09097},
      archivePrefix={arXiv},
      primaryClass={astro-ph.CO},
      url={https://arxiv.org/abs/2505.09097}, 
}

@ARTICLE{Ghara2020,
       author = {{Ghara}, R. and {Giri}, S.~K. and {Mellema}, G. and {Ciardi}, B. and {Zaroubi}, S. and {Iliev}, I.~T. and {Koopmans}, L.~V.~E. and {Chapman}, E. and {Gazagnes}, S. and {Gehlot}, B.~K. and {Ghosh}, A. and {Jeli{\'c}}, V. and {Mertens}, F.~G. and {Mondal}, R. and {Schaye}, J. and {Silva}, M.~B. and {Asad}, K.~M.~B. and {Kooistra}, R. and {Mevius}, M. and {Offringa}, A.~R. and {Pandey}, V.~N. and {Yatawatta}, S.},
        title = "{Constraining the intergalactic medium at z ≍ 9.1 using LOFAR Epoch of Reionization observations}",
      journal = {\mnras},
         year = 2020,
        month = feb,
       volume = {493},
       number = {4},
        pages = {4728-4747},
          doi = {10.1093/mnras/staa487},
archivePrefix = {arXiv},
       eprint = {2002.07195},
 primaryClass = {astro-ph.CO},
       adsurl = {https://ui.adsabs.harvard.edu/abs/2020MNRAS.493.4728G}
}

@ARTICLE{Mondal2020b,
       author = {{Mondal}, R. and {Fialkov}, A. and {Fling}, C. and {Iliev}, I.~T. and {Barkana}, R. and {Ciardi}, B. and {Mellema}, G. and {Zaroubi}, S. and {Koopmans}, L.~V.~E. and {Mertens}, F.~G. and {Gehlot}, B.~K. and {Ghara}, R. and {Ghosh}, A. and {Giri}, S.~K. and {Offringa}, A. and {Pandey}, V.~N.},
        title = "{Tight constraints on the excess radio background at z = 9.1 from LOFAR}",
      journal = {\mnras},
         year = 2020,
        month = nov,
       volume = {498},
       number = {3},
        pages = {4178-4191},
          doi = {10.1093/mnras/staa2422},
archivePrefix = {arXiv},
       eprint = {2004.00678},
 primaryClass = {astro-ph.CO},
       adsurl = {https://ui.adsabs.harvard.edu/abs/2020MNRAS.498.4178M}
}

@misc{theheracollaboration2025resultsheraphaseii,
      title={First Results from HERA Phase II}, 
      author={{The HERA Collaboration} and Zuhra Abdurashidova and Tyrone Adams and James E. Aguirre and Rushelle Baartman and Rennan Barkana and Lindsay M. Berkhout and Gianni Bernardi and Tashalee S. Billings and Bruno B. Bizarria and Judd D. Bowman and Daniela Breitman and Philip Bull and Jacob Burba and Ruby Byrne and Steven Carey and Rajorshi Sushovan Chandra and Kai-Feng Chen and Samir Choudhuri and Tyler Cox and David R. DeBoer and Eloy de Lera Acedo and Matt Dexter and Jiten Dhandha and Joshua S. Dillon and Scott Dynes and Nico Eksteen and John Ely and Aaron Ewall-Wice and Nicolas Fagnoni and Anastasia Fialkov and Steven R. Furlanetto and Kingsley Gale-Sides and Hugh Garsden and Adelie Gorce and Deepthi Gorthi and Ziyaad Halday and Bryna J. Hazelton and Jacqueline N. Hewitt and Jack Hickish and Tian Huang and Daniel C. Jacobs and Alec Josaitis and Nicholas S. Kern and Joshua Kerrigan and Piyanat Kittiwisit and Matthew Kolopanis and Adam Lanman and Paul La Plante and Adrian Liu and Yin-Zhe Ma and David H. E. MacMahon and Lourence Malan and Cresshim Malgas and Keith Malgas and Bradley Marero and Zachary E. Martinot and Lisa McBride and Andrei Mesinger and Jordan Mirocha and Nicel Mohamed-Hinds and Mathakane Molewa and Miguel F. Morales and Julian B. Muñoz and Steven G. Murray and Bojan Nikolic and Hans Nuwegeld and Aaron R. Parsons and Robert Pascua and Nipanjana Patra and Simon Pochinda and Yuxiang Qin and Eleanor Rath and Nima Razavi-Ghods and Daniel Riley and Kathryn Rosie and Mario G. Santos and Saurabh Singh and Dara Storer and Hilton Swarts and Jianrong Tan and Emilie Thélie and Pieter van Wyngaarden and Michael J. Wilensky and Peter K. G. Williams and Haoxuan Zheng},
      year={2025},
      eprint={2511.21289},
      archivePrefix={arXiv},
      primaryClass={astro-ph.CO},
      url={https://arxiv.org/abs/2511.21289}, 
}

@ARTICLE{Abdurashidova_2023,
       author = {{The HERA Collaboration} and {Abdurashidova}, Zara and {Adams}, Tyrone and {Aguirre}, James E. and {Alexander}, Paul and {Ali}, Zaki S. and {Baartman}, Rushelle and {Balfour}, Yanga and {Barkana}, Rennan and {Beardsley}, Adam P. and {Bernardi}, Gianni and {Billings}, Tashalee S. and {Bowman}, Judd D. and {Bradley}, Richard F. and {Breitman}, Daniela and {Bull}, Philip and {Burba}, Jacob and {Carey}, Steve and {Carilli}, Chris L. and {Cheng}, Carina and {Choudhuri}, Samir and {DeBoer}, David R. and {de Lera Acedo}, Eloy and {Dexter}, Matt and {Dillon}, Joshua S. and {Ely}, John and {Ewall-Wice}, Aaron and {Fagnoni}, Nicolas and {Fialkov}, Anastasia and {Fritz}, Randall and {Furlanetto}, Steven R. and {Gale-Sides}, Kingsley and {Garsden}, Hugh and {Glendenning}, Brian and {Gorce}, Ad{\'e}lie and {Gorthi}, Deepthi and {Greig}, Bradley and {Grobbelaar}, Jasper and {Halday}, Ziyaad and {Hazelton}, Bryna J. and {Heimersheim}, Stefan and {Hewitt}, Jacqueline N. and {Hickish}, Jack and {Jacobs}, Daniel C. and {Julius}, Austin and {Kern}, Nicholas S. and {Kerrigan}, Joshua and {Kittiwisit}, Piyanat and {Kohn}, Saul A. and {Kolopanis}, Matthew and {Lanman}, Adam and {La Plante}, Paul and {Lewis}, David and {Liu}, Adrian and {Loots}, Anita and {Ma}, Yin-Zhe and {MacMahon}, David H.~E. and {Malan}, Lourence and {Malgas}, Keith and {Malgas}, Cresshim and {Maree}, Matthys and {Marero}, Bradley and {Martinot}, Zachary E. and {McBride}, Lisa and {Mesinger}, Andrei and {Mirocha}, Jordan and {Molewa}, Mathakane and {Morales}, Miguel F. and {Mosiane}, Tshegofalang and {Mu{\~n}oz}, Julian B. and {Murray}, Steven G. and {Nagpal}, Vighnesh and {Neben}, Abraham R. and {Nikolic}, Bojan and {Nunhokee}, Chuneeta D. and {Nuwegeld}, Hans and {Parsons}, Aaron R. and {Pascua}, Robert and {Patra}, Nipanjana and {Pieterse}, Samantha and {Qin}, Yuxiang and {Razavi-Ghods}, Nima and {Robnett}, James and {Rosie}, Kathryn and {Santos}, Mario G. and {Sims}, Peter and {Singh}, Saurabh and {Smith}, Craig and {Swarts}, Hilton and {Tan}, Jianrong and {Thyagarajan}, Nithyanandan and {Wilensky}, Michael J. and {Williams}, Peter K.~G. and {van Wyngaarden}, Pieter and {Zheng}, Haoxuan},
        title = "{Improved Constraints on the 21 cm EoR Power Spectrum and the X-Ray Heating of the IGM with HERA Phase I Observations}",
      journal = {\apj},
         year = 2023,
        month = mar,
       volume = {945},
       number = {2},
          eid = {124},
        pages = {124},
          doi = {10.3847/1538-4357/acaf50},
archivePrefix = {arXiv},
       eprint = {2210.04912},
 primaryClass = {astro-ph.CO},
       adsurl = {https://ui.adsabs.harvard.edu/abs/2023ApJ...945..124H}
}

@article{Barry_2019,
   title={Improving the Epoch of Reionization Power Spectrum Results from Murchison Widefield Array Season 1 Observations},
   volume={884},
   ISSN={1538-4357},
   url={http://dx.doi.org/10.3847/1538-4357/ab40a8},
   DOI={10.3847/1538-4357/ab40a8},
   number={1},
   journal={The Astrophysical Journal},
   publisher={American Astronomical Society},
   author={Barry, N. and Wilensky, M. and Trott, C. M. and Pindor, B. and Beardsley, A. P. and Hazelton, B. J. and Sullivan, I. S. and Morales, M. F. and Pober, J. C. and Line, J. and Greig, B. and Byrne, R. and Lanman, A. and Li, W. and Jordan, C. H. and Joseph, R. C. and McKinley, B. and Rahimi, M. and Yoshiura, S. and Bowman, J. D. and Gaensler, B. M. and Hewitt, J. N. and Jacobs, D. C. and Mitchell, D. A. and Udaya Shankar, N. and Sethi, S. K. and Subrahmanyan, R. and Tingay, S. J. and Webster, R. L. and Wyithe, J. S. B.},
   year={2019},
   month=Oct, pages={1} }

@article{Li_2019,
   title={First Season MWA Phase II Epoch of Reionization Power Spectrum Results at Redshift 7},
   volume={887},
   ISSN={1538-4357},
   url={http://dx.doi.org/10.3847/1538-4357/ab55e4},
   DOI={10.3847/1538-4357/ab55e4},
   number={2},
   journal={The Astrophysical Journal},
   publisher={American Astronomical Society},
   author={Li, W. and Pober, J. C. and Barry, N. and Hazelton, B. J. and Morales, M. F. and Trott, C. M. and Lanman, A. and Wilensky, M. and Sullivan, I. and Beardsley, A. P. and Booler, T. and Bowman, J. D. and Byrne, R. and Crosse, B. and Emrich, D. and Franzen, T. M. O. and Hasegawa, K. and Horsley, L. and Johnston-Hollitt, M. and Jacobs, D. C. and Jordan, C. H. and Joseph, R. C. and Kaneuji, T. and Kaplan, D. L. and Kenney, D. and Kubota, K. and Line, J. and Lynch, C. and McKinley, B. and Mitchell, D. A. and Murray, S. and Pallot, D. and Pindor, B. and Rahimi, M. and Riding, J. and Sleap, G. and Steele, K. and Takahashi, K. and Tingay, S. J. and Walker, M. and Wayth, R. B. and Webster, R. L. and Williams, A. and Wu, C. and Wyithe, J. S. B. and Yoshiura, S. and Zheng, Q.},
   year={2019},
   month=Dec, pages={141} }

@article{Trott_2020,
   title={Deep multiredshift limits on Epoch of Reionization 21 cm power spectra from four seasons of Murchison Widefield Array observations},
   volume={493},
   ISSN={1365-2966},
   url={http://dx.doi.org/10.1093/mnras/staa414},
   DOI={10.1093/mnras/staa414},
   number={4},
   journal={Monthly Notices of the Royal Astronomical Society},
   publisher={Oxford University Press (OUP)},
   author={Trott, Cathryn M and Jordan, C H and Midgley, S and Barry, N and Greig, B and Pindor, B and Cook, J H and Sleap, G and Tingay, S J and Ung, D and Hancock, P and Williams, A and Bowman, J and Byrne, R and Chokshi, A and Hazelton, B J and Hasegawa, K and Jacobs, D and Joseph, R C and Li, W and Line, J L B and Lynch, C and McKinley, B and Mitchell, D A and Morales, M F and Ouchi, M and Pober, J C and Rahimi, M and Takahashi, K and Wayth, R B and Webster, R L and Wilensky, M and Wyithe, J S B and Yoshiura, S and Zhang, Z and Zheng, Q},
   year={2020},
   month=Feb, pages={4711–4727} }

@article{Mertens_2020,
   title={Improved upper limits on the 21-cm signal power spectrum of neutral hydrogen at z ≈ 9.1 from LOFAR},
   volume={493},
   ISSN={1365-2966},
   url={http://dx.doi.org/10.1093/mnras/staa327},
   DOI={10.1093/mnras/staa327},
   number={2},
   journal={Monthly Notices of the Royal Astronomical Society},
   publisher={Oxford University Press (OUP)},
   author={Mertens, F G and Mevius, M and Koopmans, L V E and Offringa, A R and Mellema, G and Zaroubi, S and Brentjens, M A and Gan, H and Gehlot, B K and Pandey, V N and Sardarabadi, A M and Vedantham, H K and Yatawatta, S and Asad, K M B and Ciardi, B and Chapman, E and Gazagnes, S and Ghara, R and Ghosh, A and Giri, S K and Iliev, I T and Jelić, V and Kooistra, R and Mondal, R and Schaye, J and Silva, M B},
   year={2020},
   month=Feb, pages={1662–1685} }

@article{Mertens_2025,
   title={Deeper multi-redshift upper limits on the epoch of reionisation 21-cm signal power spectrum from LOFAR between <i>z</i>=8.3 and <i>z</i>=10.1},
   volume={698},
   ISSN={1432-0746},
   url={http://dx.doi.org/10.1051/0004-6361/202554158},
   DOI={10.1051/0004-6361/202554158},
   journal={Astronomy &amp; Astrophysics},
   publisher={EDP Sciences},
   author={Mertens, F. G. and Mevius, M. and Koopmans, L. V. E. and Offringa, A. R. and Zaroubi, S. and Acharya, A. and Brackenhoff, S. A. and Ceccotti, E. and Chapman, E. and Chege, K. and Ciardi, B. and Ghara, R. and Ghosh, S. and Giri, S. K. and Hothi, I. and Höfer, C. and Iliev, I. T. and Jelić, V. and Ma, Q. and Mellema, G. and Munshi, S. and Pandey, V. N. and Yatawatta, S.},
   year={2025},
   month=June, pages={A186} }

@article{Planck2020,
	author = {{Planck Collaboration} and {Aghanim, N.} and {Akrami, Y.} and {Ashdown, M.} and {Aumont, J.} and {Baccigalupi, C.} and {Ballardini, M.} and {Banday, A. J.} and {Barreiro, R. B.} and {Bartolo, N.} and {Basak, S.} and {Battye, R.} and {Benabed, K.} and {Bernard, J.-P.} and {Bersanelli, M.} and {Bielewicz, P.} and {Bock, J. J.} and {Bond, J. R.} and {Borrill, J.} and {Bouchet, F. R.} and {Boulanger, F.} and {Bucher, M.} and {Burigana, C.} and {Butler, R. C.} and {Calabrese, E.} and {Cardoso, J.-F.} and {Carron, J.} and {Challinor, A.} and {Chiang, H. C.} and {Chluba, J.} and {Colombo, L. P. L.} and {Combet, C.} and {Contreras, D.} and {Crill, B. P.} and {Cuttaia, F.} and {de Bernardis, P.} and {de Zotti, G.} and {Delabrouille, J.} and {Delouis, J.-M.} and {Di Valentino, E.} and {Diego, J. M.} and {Dor\'e, O.} and {Douspis, M.} and {Ducout, A.} and {Dupac, X.} and {Dusini, S.} and {Efstathiou, G.} and {Elsner, F.} and {En\ss{}lin, T. A.} and {Eriksen, H. K.} and {Fantaye, Y.} and {Farhang, M.} and {Fergusson, J.} and {Fernandez-Cobos, R.} and {Finelli, F.} and {Forastieri, F.} and {Frailis, M.} and {Fraisse, A. A.} and {Franceschi, E.} and {Frolov, A.} and {Galeotta, S.} and {Galli, S.} and {Ganga, K.} and {G\'enova-Santos, R. T.} and {Gerbino, M.} and {Ghosh, T.} and {Gonz\'alez-Nuevo, J.} and {G\'orski, K. M.} and {Gratton, S.} and {Gruppuso, A.} and {Gudmundsson, J. E.} and {Hamann, J.} and {Handley, W.} and {Hansen, F. K.} and {Herranz, D.} and {Hildebrandt, S. R.} and {Hivon, E.} and {Huang, Z.} and {Jaffe, A. H.} and {Jones, W. C.} and {Karakci, A.} and {Keih\"anen, E.} and {Keskitalo, R.} and {Kiiveri, K.} and {Kim, J.} and {Kisner, T. S.} and {Knox, L.} and {Krachmalnicoff, N.} and {Kunz, M.} and {Kurki-Suonio, H.} and {Lagache, G.} and {Lamarre, J.-M.} and {Lasenby, A.} and {Lattanzi, M.} and {Lawrence, C. R.} and {Le Jeune, M.} and {Lemos, P.} and {Lesgourgues, J.} and {Levrier, F.} and {Lewis, A.} and {Liguori, M.} and {Lilje, P. B.} and {Lilley, M.} and {Lindholm, V.} and {L\'opez-Caniego, M.} and {Lubin, P. M.} and {Ma, Y.-Z.} and {Mac\'{\i}as-P\'erez, J. F.} and {Maggio, G.} and {Maino, D.} and {Mandolesi, N.} and {Mangilli, A.} and {Marcos-Caballero, A.} and {Maris, M.} and {Martin, P. G.} and {Martinelli, M.} and {Mart\'{\i}nez-Gonz\'alez, E.} and {Matarrese, S.} and {Mauri, N.} and {McEwen, J. D.} and {Meinhold, P. R.} and {Melchiorri, A.} and {Mennella, A.} and {Migliaccio, M.} and {Millea, M.} and {Mitra, S.} and {Miville-Desch\^enes, M.-A.} and {Molinari, D.} and {Montier, L.} and {Morgante, G.} and {Moss, A.} and {Natoli, P.} and {N\o{}rgaard-Nielsen, H. U.} and {Pagano, L.} and {Paoletti, D.} and {Partridge, B.} and {Patanchon, G.} and {Peiris, H. V.} and {Perrotta, F.} and {Pettorino, V.} and {Piacentini, F.} and {Polastri, L.} and {Polenta, G.} and {Puget, J.-L.} and {Rachen, J. P.} and {Reinecke, M.} and {Remazeilles, M.} and {Renzi, A.} and {Rocha, G.} and {Rosset, C.} and {Roudier, G.} and {Rubi\~no-Mart\'{\i}n, J. A.} and {Ruiz-Granados, B.} and {Salvati, L.} and {Sandri, M.} and {Savelainen, M.} and {Scott, D.} and {Shellard, E. P. S.} and {Sirignano, C.} and {Sirri, G.} and {Spencer, L. D.} and {Sunyaev, R.} and {Suur-Uski, A.-S.} and {Tauber, J. A.} and {Tavagnacco, D.} and {Tenti, M.} and {Toffolatti, L.} and {Tomasi, M.} and {Trombetti, T.} and {Valenziano, L.} and {Valiviita, J.} and {Van Tent, B.} and {Vibert, L.} and {Vielva, P.} and {Villa, F.} and {Vittorio, N.} and {Wandelt, B. D.} and {Wehus, I. K.} and {White, M.} and {White, S. D. M.} and {Zacchei, A.} and {Zonca, A.}},
	title = {Planck 2018 results - VI. Cosmological parameters},
	DOI= "10.1051/0004-6361/201833910",
	url= "https://doi.org/10.1051/0004-6361/201833910",
	journal = {A\&A},
	year = 2020,
	volume = 641,
	pages = "A6",
}

@article{Becker2001,
doi = {10.1086/324231},
url = {https://doi.org/10.1086/324231},
year = {2001},
month = {dec},
publisher = {},
volume = {122},
number = {6},
pages = {2850},
author = {Becker, Robert H. and Fan, Xiaohui and White, Richard L. and Strauss, Michael A. and Narayanan, Vijay K. and Lupton, Robert H. and Gunn, James E. and Annis, James and Bahcall, Neta A. and Brinkmann, J. and Connolly, A. J. and Csabai, István and Czarapata, Paul C. and Doi, Mamoru and Heckman, Timothy M. and Hennessy, G. S. and Ivezić, Željko and Knapp, G. R. and Lamb, Don Q. and McKay, Timothy A. and Munn, Jeffrey A. and Nash, Thomas and Nichol, Robert and Pier, Jeffrey R. and Richards, Gordon T. and Schneider, Donald P. and Stoughton, Chris and Szalay, Alexander S. and Thakar, Aniruddha R. and York, D. G.},
title = {Evidence for Reionization at
z ∼ 6: Detection of a
Gunn-Peterson Trough in a
z = 6.28
Quasar* **},
journal = {The Astronomical Journal}
}

@ARTICLE{Furlanetto2006,
       author = {{Furlanetto}, Steven R. and {Oh}, S. Peng and {Briggs}, Frank H.},
        title = "{Cosmology at low frequencies: The 21 cm transition and the high-redshift Universe}",
      journal = {\physrep},
         year = 2006,
        month = oct,
       volume = {433},
       number = {4-6},
        pages = {181-301},
          doi = {10.1016/j.physrep.2006.08.002},
archivePrefix = {arXiv},
       eprint = {astro-ph/0608032},
 primaryClass = {astro-ph},
       adsurl = {https://ui.adsabs.harvard.edu/abs/2006PhR...433..181F}
}

@ARTICLE{Pritchard2012,
       author = {{Pritchard}, Jonathan R. and {Loeb}, Abraham},
        title = "{21 cm cosmology in the 21st century}",
      journal = {Reports on Progress in Physics},
         year = 2012,
        month = aug,
       volume = {75},
       number = {8},
          eid = {086901},
        pages = {086901},
          doi = {10.1088/0034-4885/75/8/086901},
archivePrefix = {arXiv},
       eprint = {1109.6012},
 primaryClass = {astro-ph.CO},
       adsurl = {https://ui.adsabs.harvard.edu/abs/2012RPPh...75h6901P}
}

@ARTICLE{Morales2010,
       author = {{Morales}, Miguel F. and {Wyithe}, J. Stuart B.},
        title = "{Reionization and Cosmology with 21-cm Fluctuations}",
      journal = {\araa},
         year = 2010,
        month = sep,
       volume = {48},
        pages = {127-171},
          doi = {10.1146/annurev-astro-081309-130936},
archivePrefix = {arXiv},
       eprint = {0910.3010},
 primaryClass = {astro-ph.CO},
       adsurl = {https://ui.adsabs.harvard.edu/abs/2010ARA&A..48..127M}
}

@article{Mondal2022,
    author = {Mondal, Rajesh and Mellema, Garrelt and Murray, Steven G and Greig, Bradley},
    title = {The multifrequency angular power spectrum in parameter studies of the cosmic 21-cm signal},
    journal = {Monthly Notices of the Royal Astronomical Society: Letters},
    volume = {514},
    number = {1},
    pages = {L31-L35},
    year = {2022},
    month = {05},
    issn = {1745-3925},
    doi = {10.1093/mnrasl/slac053},
    url = {https://doi.org/10.1093/mnrasl/slac053},
    eprint = {https://academic.oup.com/mnrasl/article-pdf/514/1/L31/54615242/slac053.pdf},
}

@article{Shaw2020,
    author = {Shaw, Abinash Kumar and Bharadwaj, Somnath and Mondal, Rajesh},
    title = {The impact of non-Gaussianity on the Epoch of Reionization parameter forecast using 21-cm power-spectrum measurements},
    journal = {Monthly Notices of the Royal Astronomical Society},
    volume = {498},
    number = {1},
    pages = {1480-1495},
    year = {2020},
    month = {07},
    issn = {0035-8711},
    doi = {10.1093/mnras/staa2090},
    url = {https://doi.org/10.1093/mnras/staa2090},
    eprint = {https://academic.oup.com/mnras/article-pdf/498/1/1480/33755046/staa2090.pdf},
}

@ARTICLE{Bharadwaj2004,
       author = {{Bharadwaj}, Somnath and {Ali}, Sk. Saiyad},
        title = "{The cosmic microwave background radiation fluctuations from HI perturbations prior to reionization}",
      journal = {\mnras},
         year = 2004,
        month = jul,
       volume = {352},
       number = {1},
        pages = {142-146},
          doi = {10.1111/j.1365-2966.2004.07907.x},
archivePrefix = {arXiv},
       eprint = {astro-ph/0401206},
 primaryClass = {astro-ph},
       adsurl = {https://ui.adsabs.harvard.edu/abs/2004MNRAS.352..142B}
}

@article{Bharadwaj2020,
    author = {Bharadwaj, Somnath and Mazumdar, Arindam and Sarkar, Debanjan},
    title = {Quantifying the redshift space distortion of the bispectrum I: primordial non-Gaussianity},
    journal = {Monthly Notices of the Royal Astronomical Society},
    volume = {493},
    number = {1},
    pages = {594-602},
    year = {2020},
    month = {02},
    issn = {0035-8711},
    doi = {10.1093/mnras/staa279},
    url = {https://doi.org/10.1093/mnras/staa279},
    eprint = {https://academic.oup.com/mnras/article-pdf/493/1/594/32513251/staa279.pdf},
}

@ARTICLE{Iliev2006,
       author = {{Iliev}, I.~T. and {Mellema}, G. and {Pen}, U.-L. and {Merz}, H. and {Shapiro}, P.~R. and {Alvarez}, M.~A.},
        title = "{Simulating cosmic reionization at large scales - I. The geometry of reionization}",
      journal = {\mnras},
         year = 2006,
        month = jul,
       volume = {369},
       number = {4},
        pages = {1625-1638},
          doi = {10.1111/j.1365-2966.2006.10502.x},
archivePrefix = {arXiv},
       eprint = {astro-ph/0512187},
 primaryClass = {astro-ph},
       adsurl = {https://ui.adsabs.harvard.edu/abs/2006MNRAS.369.1625I}
}

@ARTICLE{Majumdar2018,
       author = {{Majumdar}, Suman and {Pritchard}, Jonathan R. and {Mondal}, Rajesh and {Watkinson}, Catherine A. and {Bharadwaj}, Somnath and {Mellema}, Garrelt},
        title = "{Quantifying the non-Gaussianity in the EoR 21-cm signal through bispectrum}",
      journal = {\mnras},
         year = 2018,
        month = may,
       volume = {476},
       number = {3},
        pages = {4007-4024},
          doi = {10.1093/mnras/sty535},
archivePrefix = {arXiv},
       eprint = {1708.08458},
 primaryClass = {astro-ph.CO},
       adsurl = {https://ui.adsabs.harvard.edu/abs/2018MNRAS.476.4007M}
}

@ARTICLE{Watkinson2017,
       author = {{Watkinson}, Catherine A. and {Majumdar}, Suman and {Pritchard}, Jonathan R. and {Mondal}, Rajesh},
        title = "{A fast estimator for the bispectrum and beyond - a practical method for measuring non-Gaussianity in 21-cm maps}",
      journal = {\mnras},
         year = 2017,
        month = dec,
       volume = {472},
       number = {2},
        pages = {2436-2446},
          doi = {10.1093/mnras/stx2130},
archivePrefix = {arXiv},
       eprint = {1705.06284},
 primaryClass = {astro-ph.CO},
       adsurl = {https://ui.adsabs.harvard.edu/abs/2017MNRAS.472.2436W}
}

@ARTICLE{Mondal2020,
       author = {{Mondal}, Rajesh and {Shaw}, Abinash Kumar and {Iliev}, Ilian T. and {Bharadwaj}, Somnath and {Datta}, Kanan K. and {Majumdar}, Suman and {Sarkar}, Anjan K. and {Dixon}, Keri L.},
        title = "{Predictions for measuring the 21-cm multifrequency angular power spectrum using SKA-Low}",
      journal = {\mnras},
         year = 2020,
        month = may,
       volume = {494},
       number = {3},
        pages = {4043-4056},
          doi = {10.1093/mnras/staa1026},
archivePrefix = {arXiv},
       eprint = {1910.05196},
 primaryClass = {astro-ph.CO},
       adsurl = {https://ui.adsabs.harvard.edu/abs/2020MNRAS.494.4043M}
}

@ARTICLE{Shaw2019,
       author = {{Shaw}, Abinash Kumar and {Bharadwaj}, Somnath and {Mondal}, Rajesh},
        title = "{The impact of non-Gaussianity on the error covariance for observations of the Epoch of Reionization 21-cm power spectrum}",
      journal = {\mnras},
         year = 2019,
        month = aug,
       volume = {487},
       number = {4},
        pages = {4951-4964},
          doi = {10.1093/mnras/stz1561},
archivePrefix = {arXiv},
       eprint = {1902.08706},
 primaryClass = {astro-ph.CO},
       adsurl = {https://ui.adsabs.harvard.edu/abs/2019MNRAS.487.4951S}
}

@ARTICLE{Mesinger2011,
       author = {{Mesinger}, Andrei and {Furlanetto}, Steven and {Cen}, Renyue},
        title = "{21CMFAST: a fast, seminumerical simulation of the high-redshift 21-cm signal}",
      journal = {\mnras},
         year = 2011,
        month = feb,
       volume = {411},
       number = {2},
        pages = {955-972},
          doi = {10.1111/j.1365-2966.2010.17731.x},
archivePrefix = {arXiv},
       eprint = {1003.3878},
 primaryClass = {astro-ph.CO},
       adsurl = {https://ui.adsabs.harvard.edu/abs/2011MNRAS.411..955M}
}

@article{Tiwari2022,
   title={Improving constraints on the reionization parameters using 21-cm bispectrum},
   volume={2022},
   ISSN={1475-7516},
   url={http://dx.doi.org/10.1088/1475-7516/2022/04/045},
   DOI={10.1088/1475-7516/2022/04/045},
   number={04},
   journal={Journal of Cosmology and Astroparticle Physics},
   publisher={IOP Publishing},
   author={Tiwari, Himanshu and Shaw, Abinash Kumar and Majumdar, Suman and Kamran, Mohd and Choudhury, Madhurima},
   year={2022},
   month=apr, pages={045} }

@ARTICLE{Mondal2021,
       author = {{Mondal}, Rajesh and {Mellema}, Garrelt and {Shaw}, Abinash Kumar and {Kamran}, Mohd and {Majumdar}, Suman},
        title = "{The Epoch of Reionization 21-cm bispectrum: the impact of light-cone effects and detectability}",
      journal = {\mnras},
         year = 2021,
        month = dec,
       volume = {508},
       number = {3},
        pages = {3848-3859},
          doi = {10.1093/mnras/stab2900},
archivePrefix = {arXiv},
       eprint = {2107.02668},
 primaryClass = {astro-ph.CO},
       adsurl = {https://ui.adsabs.harvard.edu/abs/2021MNRAS.508.3848M}
}

@ARTICLE{Mondal2015,
       author = {{Mondal}, R. and {Bharadwaj}, S. and {Majumdar}, S. and {Bera}, A. and {Acharyya}, A.},
        title = "{The effect of non-Gaussianity on error predictions for the Epoch of Reionization (EoR) 21-cm power spectrum.}",
      journal = {\mnras},
         year = 2015,
        month = apr,
       volume = {449},
        pages = {L41-L45},
          doi = {10.1093/mnrasl/slv015},
archivePrefix = {arXiv},
       eprint = {1409.4420},
 primaryClass = {astro-ph.CO},
       adsurl = {https://ui.adsabs.harvard.edu/abs/2015MNRAS.449L..41M}
}

@ARTICLE{Greig2015,
       author = {{Greig}, Bradley and {Mesinger}, Andrei},
        title = "{21CMMC: an MCMC analysis tool enabling astrophysical parameter studies of the cosmic 21 cm signal}",
      journal = {\mnras},
         year = 2015,
        month = jun,
       volume = {449},
       number = {4},
        pages = {4246-4263},
          doi = {10.1093/mnras/stv571},
archivePrefix = {arXiv},
       eprint = {1501.06576},
 primaryClass = {astro-ph.CO},
       adsurl = {https://ui.adsabs.harvard.edu/abs/2015MNRAS.449.4246G}
}

@software{Foreman-Mackey2013,
       author = {{Foreman-Mackey}, Daniel and {Conley}, Alex and {Meierjurgen Farr}, Will and {Hogg}, David W. and {Lang}, Dustin and {Marshall}, Phil and {Price-Whelan}, Adrian and {Sanders}, Jeremy and {Zuntz}, Joe},
        title = "{emcee: The MCMC Hammer}",
 howpublished = {Astrophysics Source Code Library, record ascl:1303.002},
         year = 2013,
        month = mar,
          eid = {ascl:1303.002},
archivePrefix = {ascl},
       eprint = {1303.002},
       adsurl = {https://ui.adsabs.harvard.edu/abs/2013ascl.soft03002F}
}

@ARTICLE{Kamran2021,
       author = {{Kamran}, Mohd and {Ghara}, Raghunath and {Majumdar}, Suman and {Mondal}, Rajesh and {Mellema}, Garrelt and {Bharadwaj}, Somnath and {Pritchard}, Jonathan R. and {Iliev}, Ilian T.},
        title = "{Redshifted 21-cm bispectrum - II. Impact of the spin temperature fluctuations and redshift space distortions on the signal from the Cosmic Dawn}",
      journal = {\mnras},
         year = 2021,
        month = apr,
       volume = {502},
       number = {3},
        pages = {3800-3813},
          doi = {10.1093/mnras/stab216},
archivePrefix = {arXiv},
       eprint = {2012.11616},
 primaryClass = {astro-ph.CO},
       adsurl = {https://ui.adsabs.harvard.edu/abs/2021MNRAS.502.3800K}
}

@ARTICLE{Zahn2011,
       author = {{Zahn}, Oliver and {Mesinger}, Andrei and {McQuinn}, Matthew and {Trac}, Hy and {Cen}, Renyue and {Hernquist}, Lars E.},
        title = "{Comparison of reionization models: radiative transfer simulations and approximate, seminumeric models}",
      journal = {\mnras},
         year = 2011,
        month = jun,
       volume = {414},
       number = {1},
        pages = {727-738},
          doi = {10.1111/j.1365-2966.2011.18439.x},
archivePrefix = {arXiv},
       eprint = {1003.3455},
 primaryClass = {astro-ph.CO},
       adsurl = {https://ui.adsabs.harvard.edu/abs/2011MNRAS.414..727Z}
}

@ARTICLE{Majumdar2014,
       author = {{Majumdar}, Suman and {Mellema}, Garrelt and {Datta}, Kanan K. and {Jensen}, Hannes and {Choudhury}, T. Roy and {Bharadwaj}, Somnath and {Friedrich}, Martina M.},
        title = "{On the use of seminumerical simulations in predicting the 21-cm signal from the epoch of reionization}",
      journal = {\mnras},
         year = 2014,
        month = oct,
       volume = {443},
       number = {4},
        pages = {2843-2861},
          doi = {10.1093/mnras/stu1342},
archivePrefix = {arXiv},
       eprint = {1403.0941},
 primaryClass = {astro-ph.CO},
       adsurl = {https://ui.adsabs.harvard.edu/abs/2014MNRAS.443.2843M}
}

@ARTICLE{Trott2016,
       author = {{Trott}, C.~M. and {Pindor}, B. and {Procopio}, P. and {Wayth}, R.~B. and {Mitchell}, D.~A. and {McKinley}, B. and {Tingay}, S.~J. and {Barry}, N. and {Beardsley}, A.~P. and {Bernardi}, G. and {Bowman}, Judd D. and {Briggs}, F. and {Cappallo}, R.~J. and {Carroll}, P. and {de Oliveira-Costa}, A. and {Dillon}, Joshua S. and {Ewall-Wice}, A. and {Feng}, L. and {Greenhill}, L.~J. and {Hazelton}, B.~J. and {Hewitt}, J.~N. and {Hurley-Walker}, N. and {Johnston-Hollitt}, M. and {Jacobs}, Daniel C. and {Kaplan}, D.~L. and {Kim}, H.~S. and {Lenc}, E. and {Line}, J. and {Loeb}, A. and {Lonsdale}, C.~J. and {Morales}, M.~F. and {Morgan}, E. and {Neben}, A.~R. and {Thyagarajan}, Nithyanandan and {Oberoi}, D. and {Offringa}, A.~R. and {Ord}, S.~M. and {Paul}, S. and {Pober}, J.~C. and {Prabu}, T. and {Riding}, J. and {Udaya Shankar}, N. and {Sethi}, Shiv K. and {Srivani}, K.~S. and {Subrahmanyan}, R. and {Sullivan}, I.~S. and {Tegmark}, M. and {Webster}, R.~L. and {Williams}, A. and {Williams}, C.~L. and {Wu}, C. and {Wyithe}, J.~S.~B.},
        title = "{CHIPS: The Cosmological H I Power Spectrum Estimator}",
      journal = {\apj},
         year = 2016,
        month = feb,
       volume = {818},
       number = {2},
          eid = {139},
        pages = {139},
          doi = {10.3847/0004-637X/818/2/139},
archivePrefix = {arXiv},
       eprint = {1601.02073},
 primaryClass = {astro-ph.IM},
       adsurl = {https://ui.adsabs.harvard.edu/abs/2016ApJ...818..139T}
}

@ARTICLE{Greig2017,
       author = {{Greig}, Bradley and {Mesinger}, Andrei},
        title = "{Simultaneously constraining the astrophysics of reionization and the epoch of heating with 21CMMC}",
      journal = {\mnras},
         year = 2017,
        month = dec,
       volume = {472},
       number = {3},
        pages = {2651-2669},
          doi = {10.1093/mnras/stx2118},
       adsurl = {https://ui.adsabs.harvard.edu/abs/2017MNRAS.472.2651G}
}

@ARTICLE{Mondal2016,
       author = {{Mondal}, Rajesh and {Bharadwaj}, Somnath and {Majumdar}, Suman},
        title = "{Statistics of the epoch of reionization 21-cm signal - I. Power spectrum error-covariance}",
      journal = {\mnras},
         year = 2016,
        month = feb,
       volume = {456},
       number = {2},
        pages = {1936-1947},
          doi = {10.1093/mnras/stv2772},
archivePrefix = {arXiv},
       eprint = {1508.00896},
 primaryClass = {astro-ph.CO},
       adsurl = {https://ui.adsabs.harvard.edu/abs/2016MNRAS.456.1936M}
}

@InProceedings{Gal2016,
  title = 	 {Dropout as a Bayesian Approximation: Representing Model Uncertainty in Deep Learning},
  author = 	 {Gal, Yarin and Ghahramani, Zoubin},
  booktitle = 	 {Proceedings of The 33rd International Conference on Machine Learning},
  pages = 	 {1050--1059},
  year = 	 {2016},
  editor = 	 {Balcan, Maria Florina and Weinberger, Kilian Q.},
  volume = 	 {48},
  series = 	 {Proceedings of Machine Learning Research},
  address = 	 {New York, New York, USA},
  month = 	 {20--22 Jun},
  publisher =    {PMLR},
  url = 	 {https://proceedings.mlr.press/v48/gal16.html}
}

@article{Kennedy2002,
    author = {Kennedy, Marc C. and O'Hagan, Anthony},
    title = {Bayesian Calibration of Computer Models},
    journal = {Journal of the Royal Statistical Society Series B: Statistical Methodology},
    volume = {63},
    number = {3},
    pages = {425-464},
    year = {2002},
    month = {01},
    issn = {1369-7412},
    doi = {10.1111/1467-9868.00294},
    url = {https://doi.org/10.1111/1467-9868.00294},
    eprint = {https://academic.oup.com/jrsssb/article-pdf/63/3/425/49590547/jrsssb_63_3_425.pdf},
}

@ARTICLE{Furlanetto2004,
       author = {{Furlanetto}, Steven R. and {Zaldarriaga}, Matias and {Hernquist}, Lars},
        title = "{Statistical Probes of Reionization with 21 Centimeter Tomography}",
      journal = {\apj},
         year = 2004,
        month = sep,
       volume = {613},
       number = {1},
        pages = {16-22},
          doi = {10.1086/423028},
archivePrefix = {arXiv},
       eprint = {astro-ph/0404112},
 primaryClass = {astro-ph},
       adsurl = {https://ui.adsabs.harvard.edu/abs/2004ApJ...613...16F}
}

@ARTICLE{Mondal2017,
       author = {{Mondal}, Rajesh and {Bharadwaj}, Somnath and {Majumdar}, Suman},
        title = "{Statistics of the epoch of reionization (EoR) 21-cm signal - II. The evolution of the power-spectrum error-covariance}",
      journal = {\mnras},
         year = 2017,
        month = jan,
       volume = {464},
       number = {3},
        pages = {2992-3004},
          doi = {10.1093/mnras/stw2599},
archivePrefix = {arXiv},
       eprint = {1606.03874},
 primaryClass = {astro-ph.CO},
       adsurl = {https://ui.adsabs.harvard.edu/abs/2017MNRAS.464.2992M}
}

@ARTICLE{Bharadwaj2005,
       author = {{Bharadwaj}, Somnath and {Pandey}, Sanjay K.},
        title = "{Probing non-Gaussian features in the HI distribution at the epoch of re-ionization}",
      journal = {\mnras},
         year = 2005,
        month = apr,
       volume = {358},
       number = {3},
        pages = {968-976},
          doi = {10.1111/j.1365-2966.2005.08836.x},
archivePrefix = {arXiv},
       eprint = {astro-ph/0410581},
 primaryClass = {astro-ph},
       adsurl = {https://ui.adsabs.harvard.edu/abs/2005MNRAS.358..968B}
}

@article{Higdon01062008,
author = {Dave Higdon and James Gattiker and Brian Williams and Maria Rightley},
title = {Computer Model Calibration Using High-Dimensional Output},
journal = {Journal of the American Statistical Association},
volume = {103},
number = {482},
pages = {570--583},
year = {2008},
publisher = {Taylor \& Francis},
doi = {10.1198/016214507000000888},
URL = { 
    
        https://doi.org/10.1198/016214507000000888
    
    

},
eprint = { 
    
        https://doi.org/10.1198/016214507000000888
    
    

}

}

@misc{tensorflow2015-whitepaper,
title={ {TensorFlow}: Large-Scale Machine Learning on Heterogeneous Systems},
url={https://www.tensorflow.org/},
note={Software available from tensorflow.org},
author={
    Mart\'{i}n~Abadi and
    Ashish~Agarwal and
    Paul~Barham and
    Eugene~Brevdo and
    Zhifeng~Chen and
    Craig~Citro and
    Greg~S.~Corrado and
    Andy~Davis and
    Jeffrey~Dean and
    Matthieu~Devin and
    Sanjay~Ghemawat and
    Ian~Goodfellow and
    Andrew~Harp and
    Geoffrey~Irving and
    Michael~Isard and
    Yangqing Jia and
    Rafal~Jozefowicz and
    Lukasz~Kaiser and
    Manjunath~Kudlur and
    Josh~Levenberg and
    Dandelion~Man\'{e} and
    Rajat~Monga and
    Sherry~Moore and
    Derek~Murray and
    Chris~Olah and
    Mike~Schuster and
    Jonathon~Shlens and
    Benoit~Steiner and
    Ilya~Sutskever and
    Kunal~Talwar and
    Paul~Tucker and
    Vincent~Vanhoucke and
    Vijay~Vasudevan and
    Fernanda~Vi\'{e}gas and
    Oriol~Vinyals and
    Pete~Warden and
    Martin~Wattenberg and
    Martin~Wicke and
    Yuan~Yu and
    Xiaoqiang~Zheng},
  year={2015},
}

@misc{chollet2015keras,
  title={Keras},
  author={Chollet, Fran\c{c}ois and others},
  year={2015},
  howpublished={\url{https://keras.io}},
}

@inproceedings{lakshminarayanan2017,
  title={Advances in neural information processing systems},
  author={Lakshminarayanan, Balaji and Pritzel, Alexander and Blundell, Charles and Guyon, I and Luxburg, UV and Bengio, S and Wallach, H and Fergus, R and Vishwanathan, S and Garnett, R},
  booktitle={Proceedings of the 31st Conference on Neural Information Processing Systems, Long Beach, CA, USA},
  pages={4--9},
  year={2017}
}

@ARTICLE{Park2019,
       author = {{Park}, Jaehong and {Mesinger}, Andrei and {Greig}, Bradley and {Gillet}, Nicolas},
        title = "{Inferring the astrophysics of reionization and cosmic dawn from galaxy luminosity functions and the 21-cm signal}",
      journal = {\mnras},
         year = 2019,
        month = mar,
       volume = {484},
       number = {1},
        pages = {933-949},
          doi = {10.1093/mnras/stz032},
archivePrefix = {arXiv},
       eprint = {1809.08995},
 primaryClass = {astro-ph.GA},
       adsurl = {https://ui.adsabs.harvard.edu/abs/2019MNRAS.484..933P}
}

@ARTICLE{Murray2020,
       author = {{Murray}, Steven and {Greig}, Bradley and {Mesinger}, Andrei and {Mu{\~n}oz}, Julian and {Qin}, Yuxiang and {Park}, Jaehong and {Watkinson}, Catherine},
        title = "{21cmFAST v3: A Python-integrated C code for generating 3D realizations of the cosmic 21cm signal.}",
      journal = {The Journal of Open Source Software},
         year = 2020,
        month = oct,
       volume = {5},
       number = {54},
          eid = {2582},
        pages = {2582},
          doi = {10.21105/joss.02582},
archivePrefix = {arXiv},
       eprint = {2010.15121},
 primaryClass = {astro-ph.IM},
       adsurl = {https://ui.adsabs.harvard.edu/abs/2020JOSS....5.2582M}
}

@misc{Jasper2026,
      title={Mitigating Simulator Dependence in AI Parameter Inference for the Epoch of Reionization: The Importance of Simulation Diversity}, 
      author={Jasper Solt and Jonathan C. Pober and Stephen H. Bach},
      year={2026},
      eprint={2601.05229},
      archivePrefix={arXiv},
      primaryClass={astro-ph.CO},
      url={https://arxiv.org/abs/2601.05229}, 
}

@misc{Cerardi2025,
      title={Implicit inference of the reionization history with higher-order statistics of the 21-cm signal}, 
      author={Nicolas Cerardi and Sambit K. Giri and Michele Bianco and Davide Piras and Emmanuel de Salis and Massimo De Santis and Merve Selcuk-Simsek and Philipp Denzel and Kelley M. Hess and M. Carmen Toribio and Franz Kirsten and Hatem Ghorbel},
      year={2025},
      eprint={2511.11568},
      archivePrefix={arXiv},
      primaryClass={astro-ph.CO},
      url={https://arxiv.org/abs/2511.11568}, 
}

@ARTICLE{Mahida2025,
       author = {{Mahida}, Yashrajsinh and {Yadav}, Sanjay Kumar and {Majumdar}, Suman and {Noble}, Leon and {Murmu}, Chandra Shekhar and {Dasgupta}, Saswata and {Dutta}, Sohini and {Tiwari}, Himanshu and {Shaw}, Abinash Kumar},
        title = "{From ANN to BNN: Inferring reionization parameters using uncertainty-aware emulators of 21-cm summaries}",
      journal = {\jcap},
         year = 2025,
        month = dec,
       volume = {2025},
       number = {12},
          eid = {055},
        pages = {055},
          doi = {10.1088/1475-7516/2025/12/055},
archivePrefix = {arXiv},
       eprint = {2508.13261},
 primaryClass = {astro-ph.CO},
       adsurl = {https://ui.adsabs.harvard.edu/abs/2025JCAP...12..055M}
}

@ARTICLE{Noble2024,
       author = {{Noble}, Leon and {Kamran}, Mohd and {Majumdar}, Suman and {Murmu}, Chandra Shekhar and {Ghara}, Raghunath and {Mellema}, Garrelt and {Iliev}, Ilian T. and {Pritchard}, Jonathan R.},
        title = "{Impact of the Epoch of Reionization sources on the 21-cm bispectrum}",
      journal = {\jcap},
         year = 2024,
        month = oct,
       volume = {2024},
       number = {10},
          eid = {003},
        pages = {003},
          doi = {10.1088/1475-7516/2024/10/003},
archivePrefix = {arXiv},
       eprint = {2406.03118},
 primaryClass = {astro-ph.CO},
       adsurl = {https://ui.adsabs.harvard.edu/abs/2024JCAP...10..003N}
}

\appendix

\section{Error convergence diagnostic}
\label{app:variance_convergence}
To assess the stability of the error estimates, we compute the cumulative standard deviation as a function of the number of realizations $N$ for representative modes of the power spectrum and bispectrum at $z=8$. Figure~\ref{fig:variance_convergence} shows the results. For the power spectrum (panels a and b), the cumulative $\sigma$ largely stabilizes by $N \sim 40$ and remains close to the final estimate at $N=50$. For the bispectrum (panels c and d, showing squeezed-limit and equilateral configurations, respectively), the error approaches the final estimate at $N=50$ but exhibits larger residual fluctuations and does not fully stabilize.

Although the bispectrum variance has not fully converged at $N=50$, the residual sampling uncertainty is subdominant to the broader modeling uncertainties inherent in the present cross-simulation framework (Section~\ref{sec:inference}). Given the high computational cost of extending the calculation to more than 50 realizations across all redshifts and triangle configurations, we adopt 50 as a practical compromise. We therefore conclude that 50 realizations provide sufficiently stable error estimates for the purpose of this work, while larger simulation ensembles would be desirable for future analyzes.

\section{Comparison of ionization histories}
\label{app:xhi_comparison}
An important aspect of cross-simulation validation is to verify that the fiducial model is adequately covered by the training set. If the truth lies outside the range of ionization histories spanned by the {\sf 21cmFAST} training ensemble, the emulator would be forced to extrapolate, leading to biased parameter estimation. We therefore examine the distribution of $\xb$ trajectories in our 1,000 {\sf 21cmFAST} training set and compare it with the {\sf ReionYuga} fiducial ionization history.

\begin{figure}
\centering
\includegraphics[width=0.9\columnwidth]{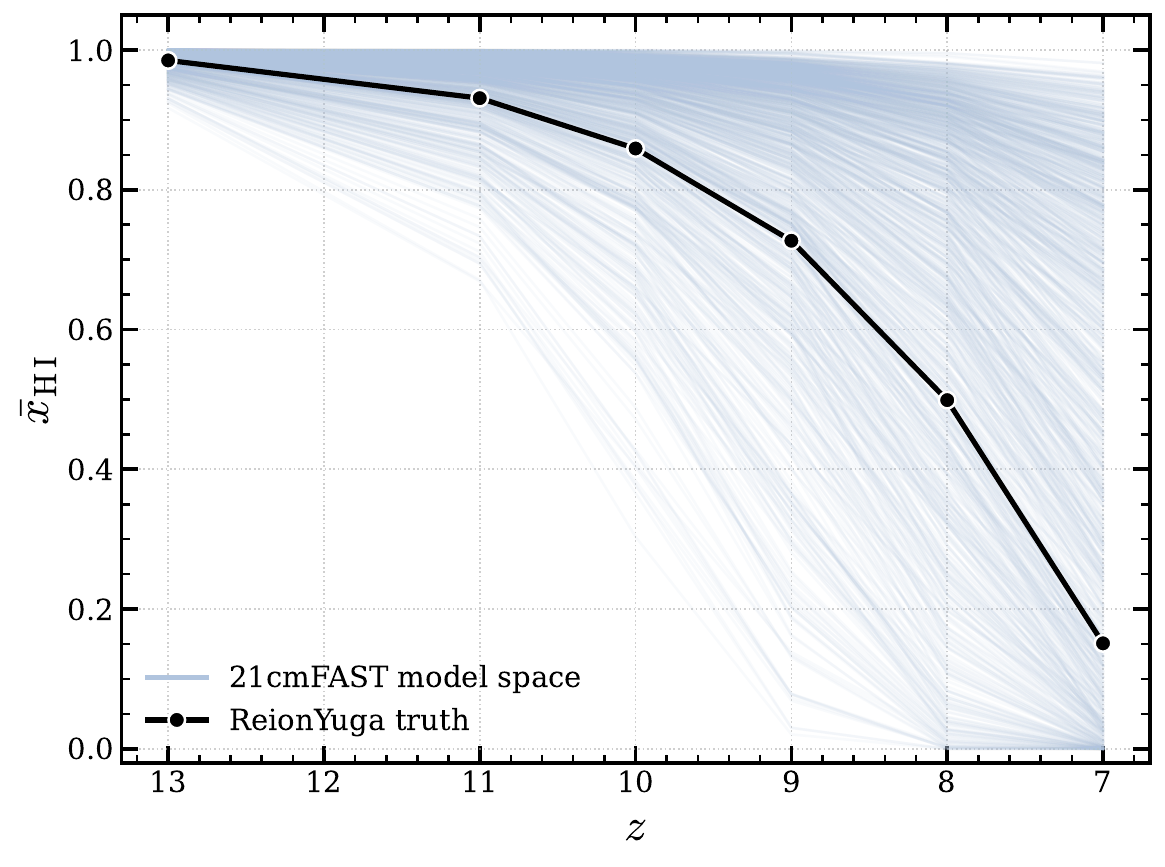}
\caption{The distribution of $\xb(z)$ histories from the {\sf 21cmFAST} training set (blue lines), overlaid with the {\sf ReionYuga} fiducial truth (solid black line with markers). The {\sf ReionYuga} history lies within the {\sf 21cmFAST} model space at all analyzed redshifts, confirming that the emulator interpolates within its training distribution throughout the inference.}
\label{fig:xhi_band}
\end{figure}

\begin{table}
\centering
\caption{Coverage of the {\sf 21cmFAST} training set at each redshift. The columns list the minimum and maximum $\xb$ values spanned by the 1000-point Latin hypercube training sample, together with the {\sf ReionYuga} truth value. The truth lies within the training bounds at all six redshifts.}
\label{tab:coverage}
\begin{tabular}{cccc}
\hline
$z$ & $\bar{x}^\mathrm{min}_\mathrm{H\,\textsc{i}}$ & $\bar{x}^\mathrm{max}_\mathrm{H\,\textsc{i}}$ & $\bar{x}^\mathrm{truth}_\mathrm{H\,\textsc{i}}$ \\
\hline
7.0  & 0.001 & 0.901 & 0.151 \\
8.0  & 0.016 & 0.950 & 0.499 \\
9.0  & 0.281 & 0.986 & 0.727 \\
10.0 & 0.568 & 0.990 & 0.859 \\
11.0 & 0.698 & 1.000 & 0.931 \\
13.0 & 0.937 & 1.000 & 0.985 \\
\hline
\end{tabular}
\end{table}

Figure~\ref{fig:xhi_band} shows all the 1,000 ionization histories from the {\sf 21cmFAST} Latin hypercube sample, overlaid with the {\sf ReionYuga} fiducial history (solid black line with markers). The {\sf 21cmFAST} ensemble spans a wide range of ionization histories, showing that the broad prior ranges adopted for the astrophysical parameters (Section~\ref{sec:inference_model}) ensure adequate coverage of the relevant model space. The {\sf ReionYuga} ionization history lies within the envelope spanned by the training set at all six redshifts, but is shifted toward somewhat later reionization than the median of the {\sf 21cmFAST} histories. Table~\ref{tab:coverage} summarizes this coverage by listing the minimum and maximum values of $\xb$ spanned by the training set together with the corresponding truth value. At all the redshifts, $\xb$ lies within the interval. Therefore, the {\sf 21cmFAST} training set, although more restricted in shape, still covers the {\sf ReionYuga} fiducial at all analyzed redshifts. The emulator, therefore, operates within its training domain and is not required to extrapolate.
\section{Bispectrum emulator performance}
\label{app:emulator_performance}

The bispectrum presents a significantly more demanding emulation challenge than the power spectrum. Its amplitude varies substantially across different triangle configurations and redshifts, spanning several orders of magnitude, and the signal can change sign. Furthermore, the mapping from $\xb$ to individual bispectrum configurations is inherently more complex and nonlinear than the corresponding power spectrum mapping, reflecting the richer phase information encoded in three-point statistics. To validate the reliability of our pointwise ensemble emulator on the configurations that dominate the likelihood, we performed several diagnostic tests on the held-out test set.

\begin{figure}
    \centering
    \includegraphics[width=\columnwidth]{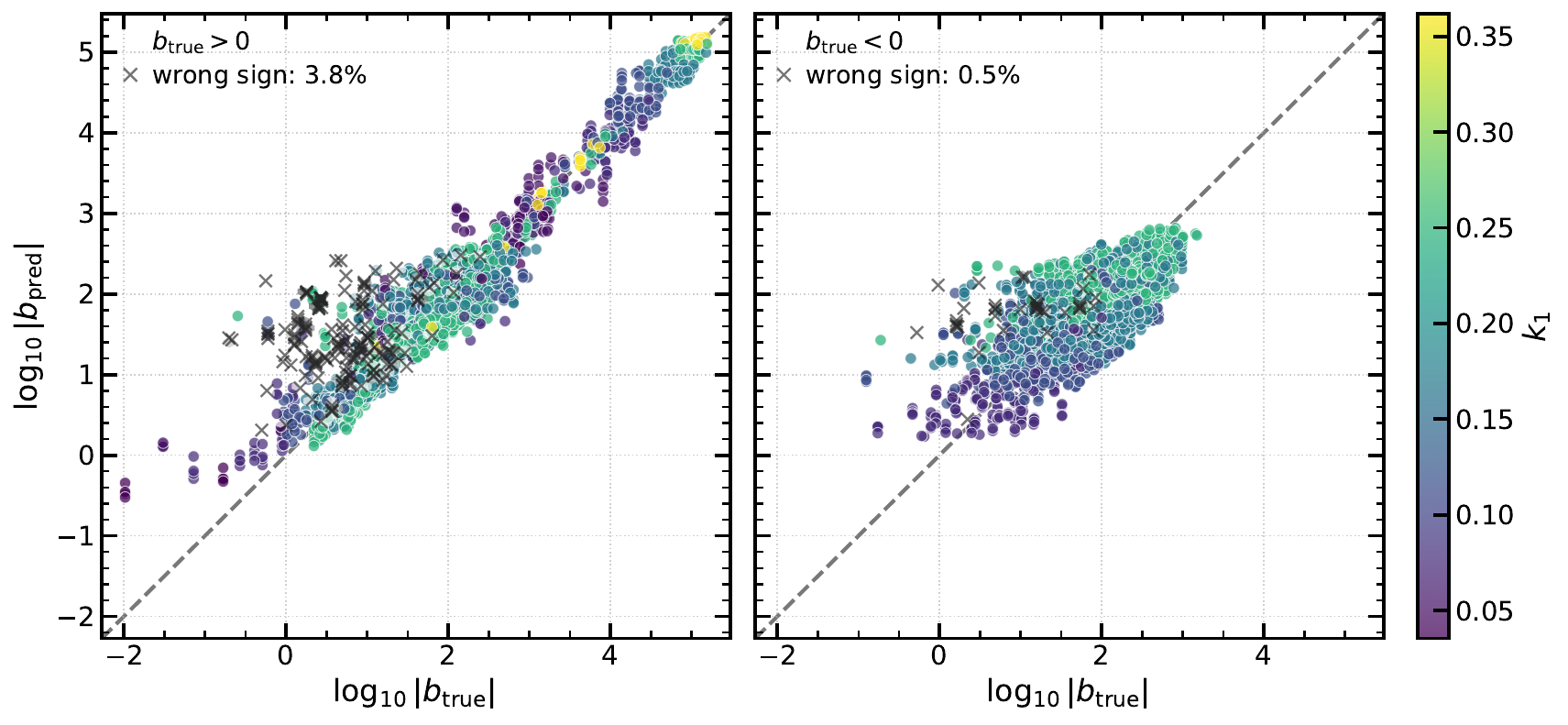}
    \caption{Predicted versus true dimensionless bispectrum amplitude in log-log scale at $z=8$, evaluated on held-out test models restricted to SNR$>1$ configurations. The left and right panels separating cases with $\tilde{b}_{\rm true}>0$ and $\tilde{b}_{\rm true}<0$, respectively. Points are coloured by $k_1$, and the dashed line denotes the 1:1 relation. The annotated wrong-sign fraction indicates the fraction of points for which the predicted bispectrum sign differs from the true sign.}
    \label{fig:frac_residual}
\end{figure}

Figure \ref{fig:frac_residual} compares the predicted and true dimensionless bispectrum amplitudes at \(z=8\) for held-out test models, using all triangle configurations retained for the likelihood analysis. The emulator reproduces the overall amplitude trend over several orders of magnitude, with the largest-amplitude configurations lying close to the one-to-one relation. The scatter is more pronounced at lower amplitudes, where the bispectrum signal is weaker and relative errors are naturally larger. The sign information is also recovered with high accuracy, with only a small fraction of wrong-sign predictions. Most of these wrong-sign cases occur in the low-amplitude regime, where modest absolute errors can move the predicted bispectrum across zero. Thus, the sign mismatches mainly reflect the difficulty of resolving weak bispectrum configurations rather than a failure to recover the dominant high-amplitude signal.

\begin{figure}
    \centering
    \includegraphics[width=0.8\columnwidth]{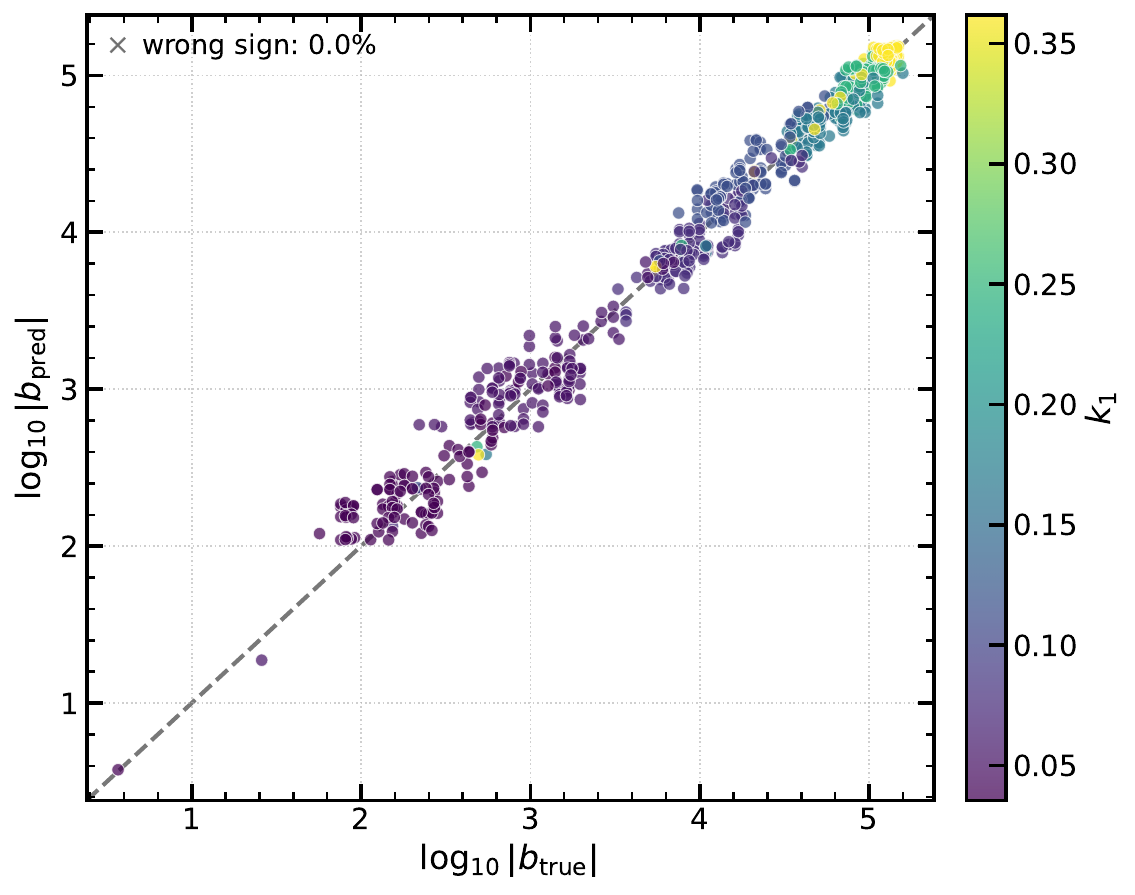}
    \caption{Predicted versus true dimensionless bispectrum amplitudes in log-log scale for squeezed-limit configurations ($n \to 1$, $\cos\theta \to 1$) at $z=8$, shown for the held-out test models restricted to SNR$>1$ configurations. The points are coloured by $k_1$. The dashed diagonal denotes perfect agreement in amplitude. Cross markers would indicate cases where the predicted bispectrum sign differs from the true sign, and the corresponding wrong-sign fraction is reported in the panel.}
    \label{fig:squeezed_scatter}
\end{figure}

Figure \ref{fig:squeezed_scatter} shows the predicted and true dimensionless bispectrum amplitudes at \(z=8\) for squeezed-limit configurations restricted to SNR\(>1\). The agreement is very tight, with the points closely following the one-to-one relation across the full amplitude range. No wrong-sign predictions are found in the held-out test set for these configurations, indicating that the emulator accurately recovers both the magnitude and the sign of the bispectrum for the high-SNR squeezed configurations used in the
likelihood analysis.

\section{Impact of calibration parameters on cross-simulation inference}
\label{app:calibration_impact}

\begin{figure*}
    \centering
    \includegraphics[width=\textwidth]{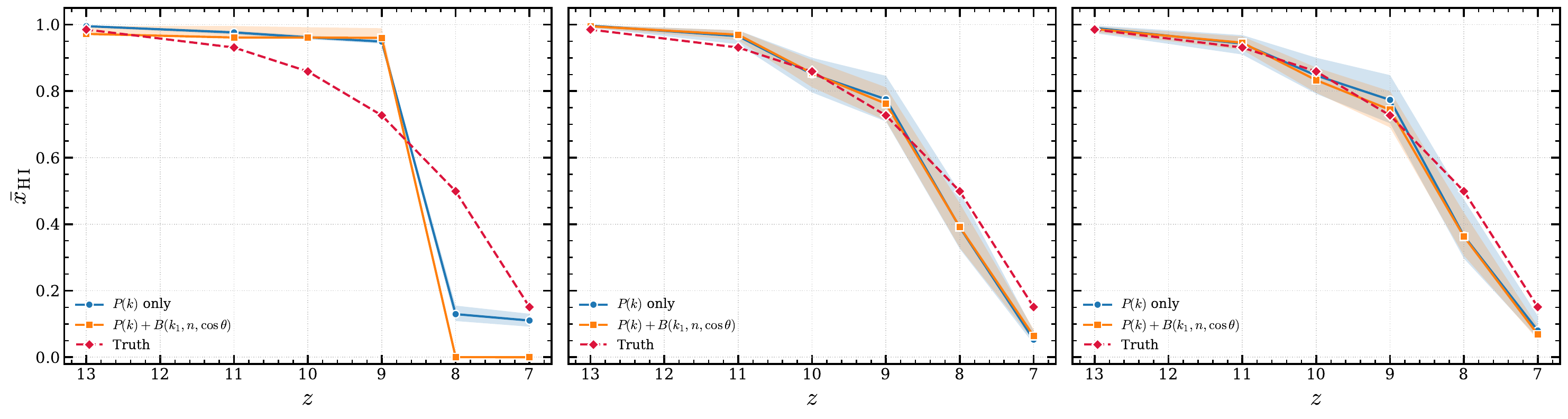}
    \caption{Comparison of inferred ionization histories for three calibration choices in the cross-simulation analysis. Blue and orange points with error bars show the inferred $\xb$ from the power spectrum alone and from the joint power spectrum--bispectrum analysis, respectively, and the red dashed line with filled diamonds shows the fiducial truth. The three panels correspond to: (1) no calibration ($A=1$, $\delta=0$), (2) redshift-independent calibration ($A=\mathrm{constant}$, $\delta=\mathrm{constant}$), and (3) redshift-dependent amplitude calibration, $A(z)=A_1(z-z_{\rm ref})+A_{0}$, with $\delta$ kept constant.}
    \label{fig:calibration_comparison}
\end{figure*}

Figure~\ref{fig:calibration_comparison} highlights the importance of the calibration parameters $A$ and $\delta_\mathrm{global}$ in the cross-simulation analysis. The left panel shows the inferred ionization histories obtained without calibration, i.e. with $A = 1$ and $\delta_\mathrm{global} = 0$. In this case, the inference fails at $z = 7$--$9$, with the posteriors collapsing to unrealistically low values of $\xb$. This occurs because the uncalibrated {\sf 21cmFAST} emulator systematically misestimates the summary-statistic amplitudes relative to the {\sf ReionYuga} mock observation, causing the sampler to compensate by moving towards highly ionized solutions.

The middle and right panels show the corresponding results with the calibration scheme of Section~\ref{sec:inference}. In both cases, the nuisance parameters remove the catastrophic failure seen without calibration and bring the inferred histories much closer to the fiducial truth. The redshift-independent calibration in the middle panel provides a substantial improvement, but residual offsets remain, particularly at intermediate redshifts. Allowing the amplitude parameter to vary with redshift in the right panel further improves the agreement, showing that the cross-simulation mismatch is not fully captured by a purely global calibration.

\bsp	
\label{lastpage}
\end{document}